\documentclass[12pt]{article}
\usepackage{amsmath,amsthm,amsfonts,amssymb}
\def\lan{\langle}
\def\ran{\rangle}

\def\Diff{{\rm Diff}}

\def\Hom{{\rm Hom}}
\def\id{{\rm id}}

\def\Vir{{\rm Vir}}

\def\a{\alpha}

\def\epsilon{\varepsilon}

\def\la{\lambda}

\def\phi{\varphi}
\def\r{\rho}
\def\s{{\sigma}}

\def\th{\theta}

\def\Om{\Omega}

\newtheorem{theorem}{Theorem}[section]
\newtheorem{lemma}[theorem]{Lemma}

\newtheorem{corollary}[theorem]{Corollary}

\newtheorem{proposition}[theorem]{Proposition}
\newtheorem{remark}[theorem]{Remark}

\def\emptyset{\varnothing}
\def\setminus{\smallsetminus}

\def\res{{\upharpoonright}}
\def\PSL{{{\rm PSL}(2,\mathbb R)}}

\def\A{{\cal A}}
\def\B{{\cal B}}
\def\CC{{\cal C}}
\def\D{{\cal D}}

\def\I{{\cal I}}
\def\K{{\cal K}}

\def\L{{\cal L}}
\def\N{{\cal A}}
\def\M{{\cal B}}
\def\E{{\cal E}}
\def\H{{\cal H}}

\def\Z{{\mathbb Z}}
\def\C{{\mathbb C}}
\def\T{{\mathbb T}}
\def\R{{\mathbb R}}

\title{{\bf Classification of Local Conformal Nets. Case $c<1$}}

\author{
{\sc Yasuyuki Kawahigashi}\footnote{Supported in part by the
Grants-in-Aid for Scientific Research, JSPS.}\\
Department of Mathematical Sciences\\
University of Tokyo, Komaba, Tokyo, 153-8914, JAPAN\\
e-mail: {\tt yasuyuki@ms.u-tokyo.ac.jp}\\
\vphantom{X}\\
{\sc Roberto Longo}\footnote{Supported in part by the Italian MIUR and 
GNAMPA-INDAM.}\\
Dipartimento di Matematica\\
Universit\`a di Roma ``Tor Vergata''\\
Via della Ricerca Scientifica 1, I-00133 Roma, ITALY\\
e-mail: {\tt longo@mat.uniroma2.it}}
\begin{document}
\date{}
\maketitle
\centerline{\sl Dedicated to Masamichi Takesaki on the occasion of his
seventieth birthday}

\begin{abstract}
We completely classify diffeomorphism covariant local nets of von 
Neumann algebras on the circle with central charge $c$ less than 1.  
The irreducible ones are in bijective correspondence with the pairs of
$A$-$D_{2n}$-$E_{6,8}$ Dynkin diagrams such that 
the difference of their Coxeter numbers is equal to 1.

We first identify the nets generated by irreducible representations of 
the Virasoro algebra for $c<1$ with certain coset nets.  Then, by 
using the classification of modular invariants for the minimal models 
by Cappelli-Itzykson-Zuber and the method of $\alpha$-induction in 
subfactor theory, we classify all local irreducible extensions of the 
Virasoro nets for $c<1$ and infer our main classification result.  
As an application, we identify in our classification list 
certain concrete coset nets studied in the literature.
\end{abstract}
\newpage

\section{Introduction}
\label{intro}
Conformal Field Theory on $S^1$ has been extensively studied in 
recent years by different methods with important motivations coming 
from various subjects of Theoretical Physics (two-dimensional critical 
phenomena, holography, \dots) and Mathematics 
(quantum groups, subfactors, topological invariants in three
dimensions, \dots).

In various approaches to the subject, it is unclear whether different 
models are to be regarded equivalent or to contain the same physical 
information.
This becomes clearer by considering the operator algebra generated 
by smeared fields
localized in a given interval $I$ of $S^1$ and take its closure 
$\A(I)$ in the weak operator topology. The relative positions of the 
various von Neumann algebras $\A(I)$,
namely the net $I\to\A(I)$, essentially encode all the structural 
information, in particular the fields can be constructed out 
of a net \cite{FJ}. 

One can describe local conformal nets by a natural set of axioms. 
The classification of such nets is certainly a well-posed problem and obviously
one of the basic ones of the subject. Note that the isomorphism class 
of a given net corresponds to the Borchers' class for the generating 
field.

Our aim in this paper is to give a first general and complete
classification of local conformal nets on $S^1$
when the central charge $c$ is less than 
1, where the central charge is the one associated with the 
representation of the Virasoro algebra (or, in physical terms, with the 
stress-energy tensor) canonically associated with 
the irreducible local conformal net, as we will explain.

Haag-Kastler nets of operator algebras have been studied in algebraic
quantum field theory for a long time (see \cite{H}, for example).
More recently, (irreducible, local) conformal nets of 
von Neumann algebras on $S^1$
have been studied, see  
\cite{BGL,C,DRL,FJ,FRS,FG,GL2,GLW,X3,X4,X5,X6,X7}.
Although a complete classification seems to be presently still
out of reach, we will make a first step 
by classifying the discrete series.

In general, it is not clear what kind of axioms we
should impose on conformal nets, beside the general ones, 
in order to obtain an interesting
mathematical structure or classification theory.
A set of conditions studied by us in \cite{KLM},
called {\sl complete rationality}, selects a basic class of nets.
Complete rationality consists of the following three requirements:
\begin{enumerate}
\item Split property.
\item Strong additivity.
\item Finiteness of the Jones index for the 2-interval inclusion.
\end{enumerate}
Properties 1 and 2 are quite general and well studied (see e.g. 
\cite{DL,GLW}). The third condition means the following.
Split the circle $S^1$ into four proper intervals and label their
interiors by $I_1, I_2, I_3, I_4$ in clockwise order.  Then,
for a local net $\A$, we have an inclusion
\[
\A(I_1)\vee \A(I_3)\subset (\A(I_2)\vee \A(I_4))' ,
\]
the ``2-interval inclusion'' of the net; its index, called the 
$\mu$-{\sl index} of $\A$, is required to be finite.

Under the assumption of complete rationality, we have proved
in \cite{KLM} that
the net has only finitely many inequivalent irreducible representations,
all have finite statistical dimensions, and the associated
braiding is non-degenerate.  That is, irreducible 
Doplicher-Haag-Roberts (DHR) endomorphisms of
the net (which basically corresponds to primary fields)
produce a modular tensor category in the sense of \cite{Tu}.
Such finiteness of the set of irreducible representations 
(``rationality'', cf. \cite{BPZ}) 
is often difficult to prove by other methods.
Furthermore, the non-degeneracy of the braiding,
also called {\sl modularity} or invertibility of the $S$-matrix,
plays an important r\^{o}le in theory of topological
invariants \cite{Tu}, particularly of Reshetikhin-Turaev type,
and is usually the hardest to prove among the axioms of modular
tensor category.  Thus our results in \cite{KLM} show that
complete rationality specifies a class of conformal nets with 
the right rational behavior.

The finiteness of the $\mu$-index may be difficult to verify directly
in concrete models as in \cite{X3}, but once this is established for 
some net, then it passes to subnets or extensions with finite index.  
Strong additivity is also often difficult to check, but recently 
one of us has proved in \cite{L4} that complete rationality 
also passes to a subnet or extension with finite index.
In this way, we now know that large classes of coset models 
\cite{X4} and orbifold models \cite{X7} are completely rational.

Now consider an irreducible local conformal net $\A$ on $S^1$. Because 
of diffeomorphism covariance, $\A$ canonically contains a subnet 
$\A_{\Vir}$
generated by a unitary projective representation of the diffeomorphism 
group of $S^1$, thus we have a representation of the Virasoro 
algebra. (In physical terms, this appears by L\"{u}scher-Mack theorem 
as Fourier modes of a chiral component of the stress-energy tensor $T$
\begin{equation*}
T(z)=\sum L_n z^{-n-2},\qquad
[L_m,L_n]=(m-n) L_{m+n} + \frac{c}{12}(m^3-m)\delta_{m,-n}.)
\end{equation*}
This representation 
decomposes into irreducible representations, all with the same 
central charge $c>0$, that is clearly an invariant for $\A$.
As is well known either $c\geq 1$ or $c$ takes a discrete set of 
values \cite{FQS}.

Our first observation is that if 
$c$ belongs to the discrete 
series, then $\A_{\Vir}$ is an irreducible subnet with finite index of 
$\A$. The classification problem for $c<1$ thus becomes the classification 
of irreducible local finite-index extensions $\A$ of the 
Virasoro nets for $c<1$. We shall show that the nets $\A_{\Vir}$ are 
completely rational if $c<1$, and so must be the original nets $\A$.

Thus, while our main result concerns nets of single factors, our main 
tool is the theory of nets of {\sl subfactors}. This is the key of our 
approach.

The outline of this paper is as follows.
We first identify the Virasoro nets with central charge less than one
and the coset net arising from the diagonal
embedding $SU(2)_{m-1}\subset SU(2)_{m-2}\times SU(2)_1$ studied in
\cite{X4}, as naturally expected from the  coset construction
of \cite{GKO}.
Then it follows from \cite{L4} that the Virasoro nets with central
charge less than 1 are completely rational.

Next we study the extensions of the Virasoro nets with central charge
less than 1.  If we have an extension, we can apply the
machinery of $\alpha$-induction, which has been introduced in \cite{LR}
and further studied in \cite{X1,X2,BE,BE4,BEK1,BEK2,BEK3}. 
This is a method producing endomorphisms of the extended net from
DHR endomorphisms of the smaller net using a braiding, but the
extended endomorphisms are not DHR endomorphisms in general.
For two irreducible DHR endomorphisms $\la,\mu$ of the smaller net,
we can make extensions $\a_\la^+,\a_\mu^-$ using positive
and negative braidings, respectively.  Then we have a non-negative
integer $Z_{\la\mu}=\dim\Hom(\a_\la^+,\a_\mu^-)$.
Recall that a completely rational net
produces a unitary representation of $SL(2,\Z)$ by \cite{R1}
and \cite{KLM} in general.
Then \cite[Corollary 5.8]{BEK1} says that 
this matrix $Z$ with non-negative integer entries and normalization
$Z_{00}=1$ is in the commutant of this unitary representation,
regardless whether the extension is local or not, and this
gives a very strong constraint on possible extensions of the Virasoro
net.  Such a matrix $Z$ is called a {\sl modular invariant} in general
and has been extensively studied in conformal
field theory.  (See \cite[Chapter 10]{DMS} for example.)
For a given unitary representation of $SL(2,\Z)$,
the number of modular invariants is always finite and often very
small, such as 1, 2, or 3, in concrete examples.  The complete classification
of modular invariants for a given representation of $SL(2,\Z)$
was first given in \cite{CIZ} for the case of the $SU(2)_k$ WZW-models
and the minimal models, and several more classification results have been
obtained by Gannon.  (See \cite{G} and references there.)

Our approach to the classification problem
of local extensions of a given net makes use of the classification of 
the modular invariants. For any local extension, we have indeed a 
modular invariant coming from the theory of $\alpha$-induction as
explained above.  For each modular invariant in the
classification list, we check the existence and uniqueness of corresponding
extensions.  In complete generality, we expect neither existence
nor uniqueness, but this approach is often powerful enough to get
a complete classification in concrete examples.
This is the case of $SU(2)_k$. (Such a classification is implicit
in \cite{BEK2}, though not explicitly stated there in this way.
See Theorem \ref{class-SU2} below.)
Also along this approach,
we obtain a complete classification of the local extensions of
the Virasoro nets with central charge less than 1 in Theorem
\ref{Vir-ext}.  By the stated canonical appearance of the Virasoro nets 
as subnets, we derive our final classification in Theorem \ref{diffeo}.
That is, our labeling of a conformal net in terms of pairs of Dynkin 
diagrams is given as follows.  For a given conformal net with
central charge $c<1$, we have a Virasoro subnet.  Then the 
$\a$-induction applied to this extension of the Virasoro net produces
a modular invariant $Z_{\la\mu}$ as above and such a matrix is
labeled with a pair of Dynkin diagrams as in \cite{CIZ}.  This
labeling gives a complete classification of such conformal nets. 

Some extensions of the Virasoro nets
in our list have been studied
or conjectured by other authors \cite{BE,X6} (they are related to the 
notion of $W$-algebra in the physical literature).  Since our classification 
is complete, it is not difficult to identify
them in our list.  This will be done in Section \ref{appl}.

Before closing this introduction we indicate possible background references 
to aid the readers, some have been already mentioned. 
Expositions of the basic structure of conformal nets 
on $S^1$ and subnets are contained in \cite{GL2} and \cite{LR}, respectively. 
Jones index theory \cite{J} is discussed in \cite{L1} in connection to 
Quantum Field Theory. Concerning modular invariants and 
$\alpha$-induction one can look at ref. \cite{BE,BEK1,BEK2}. The books
\cite{DMS,H,EK,KR} deal respectively with conformal field theory from
the physical viewpoint, algebraic quantum field theory, subfactors and 
connections with mathematical physics and infinite dimensional Lie algebras.

\section{Preliminaries}
\label{prelim}

In this section, we recall and prepare necessary results on
extensions of completely rational nets in
connection to extensions of the Virasoro nets.

\subsection{Conformal nets on $S^1$}
\label{nets}

We denote by $\I$ the family of proper intervals of $S^1$. 
A {\it net} $\A$ of von Neumann algebras on $S^1$ is a map 
$$
I\in\I\to\A(I)\subset B(\H)
$$
from $\I$ to von Neumann algebras on a fixed Hilbert space $\H$
that satisfies:
\begin{itemize}
\item[{\bf A.}] {\it Isotony}. If $I_{1}\subset I_{2}$ belong to $\I$, then
\begin{equation*}
 \A(I_{1})\subset\A(I_{2}).
\end{equation*}
\end{itemize}
The net $\A$ is called {\it local} if it satisfies:
\begin{itemize}
\item[{\bf B.}] {\it Locality}. If $I_{1},I_{2}\in\I$ and $I_1\cap 
I_2=\emptyset$ then 
\begin{equation*}
 [\A(I_{1}),\A(I_{2})]=\{0\},
 \end{equation*}
where brackets denote the commutator.
\end{itemize}
The net $\A$ is called {\it M\"{o}bius covariant} if in addition satisfies
the following properties {\bf C,D,E}:
\begin{itemize}
\item[{\bf C.}] {\it M\"{o}bius covariance}\footnote{M\"{o}bius 
covariant nets are often called {\it conformal} nets. In this paper 
however we shall reserve the term `conformal' to indicate 
diffeomorphism covariant nets.}. 
There exists a strongly 
continuous unitary representation $U$ of $\PSL$ on $\H$ such that
\begin{equation*}
 U(g)\A(I) U(g)^*\ =\ \A(gI),\quad g\in\PSL,\ I\in\I.
\end{equation*}
Here $\PSL$ acts on $S^1$ by M\" obius transformations.
\item[{\bf D.}] {\it Positivity of the energy}. The generator of the 
one-parameter rotation subgroup of $U$ (conformal Hamiltonian) is positive. 
\item[{\bf E.}] {\it Existence of the vacuum}. There exists a unit 
$U$-invariant vector $\Omega\in\H$ (vacuum vector), and $\Omega$ 
is cyclic for the von Neumann algebra $\bigvee_{I\in\I}\A(I)$.
\end{itemize}
(Here the lattice symbol $\bigvee$ denotes the von Neumann algebra 
generated.)

Let $\A$ be an irreducible M\"{o}bius covariant net. By the Reeh-Schlieder 
theorem the vacuum vector $\Om$ is cyclic and separating for each 
$\A(I)$. The Bisognano-Wichmann property then holds 
\cite{BGL,FG}: the Tomita-Takesaki modular operator $\Delta_I$ and 
conjugation $J_I$ associated with $(\A(I),\Omega)$, $I\in\I$, 
are given by 
\begin{equation}\label{BW} 
U(\Lambda_I (2\pi 
t))=\Delta_{I}^{it},\ t\in\mathbb R,\qquad U(r_I)= J_I, 
\end{equation} 
where $\Lambda_I$ is the one-parameter subgroup of $\PSL$ of special 
conformal transformations preserving $I$ and $U(r_I)$ implements a 
geometric action on $\A$ corresponding to the M\" obius reflection $r_I$ on 
$S^1$ mapping $I$ onto $I^\prime$, i.e. fixing the boundary points of 
$I$, see \cite{BGL}.

This immediately implies Haag duality (see \cite{HL,BS}): 
$$
\A(I)'=\A(I'),\quad I\in\I\ ,
$$
where $I'\equiv S^1\setminus I$.

We shall say that a M\"{o}bius covariant net $\A$ is {\it irreducible} if 
$\bigvee_{I\in\I}\A(I)=B(\H)$. Indeed $\A$ is irreducible iff
$\Om$ is the unique $U$-invariant vector (up to scalar multiples), and 
iff the local von Neumann 
algebras $\A(I)$ are factors. In this case they are III$_1$-factors 
(unless $\A(I)=\C$ identically), see \cite{GL2}.

Because of Lemma \ref{irr} below,
we may always consider irreducible nets. Hence, 
from now on, we shall make the assumption:

\begin{itemize}
\item[{\bf F.}] {\it Irreducibility}. The net $\A$ is irreducible.
\end{itemize}
Let $\Diff(S^1)$ be the group of orientation-preserving smooth 
diffeomorphisms of $S^1$. As is well known $\Diff(S^1)$ is an infinite 
dimensional Lie group whose Lie algebra is the Virasoro algebra (see 
\cite{PS,KR}).

By a {\it conformal net} (or diffeomorphism covariant 
net)  $\A$ we shall mean a M\"{o}bius covariant 
net such that the following holds:
\begin{itemize}
\item[{\bf G.}] {\it Conformal covariance}. There exists a projective unitary 
representation $U$ of $\Diff(S^1)$ on $\H$ extending the unitary 
representation of $\PSL$ such that for all $I\in\I$ we have
\begin{gather*}
 U(g)\A(I) U(g)^*\ =\ \A(gI),\quad  g\in\Diff(S^1), \\
 U(g)AU(g)^*\ =\ A,\quad A\in\A(I),\ g\in\Diff(I'),
\end{gather*}
\end{itemize}
where $\Diff(I)$ denotes the group of 
smooth diffeomorphisms $g$ of $S^1$ such that $g(t)=t$ for all $t\in I'$.

If $\A$ is a local conformal net on $S^1$ then, by Haag duality, we have 
\[
U(\Diff(I))\subset\A(I),
\]
Notice that, in general, $U(g)\Om\neq\Om$,  $g\in\Diff(S^1) $. 
Otherwise the Reeh-Schlieder theorem would be violated.
\begin{lemma}\label{irr}
Let $\A$ be a local M\"{o}bius (resp. diffeomorphism) covariant net. 
The center $Z$ of $\A(I)$ 
does not depend on the interval $I$ and $\A$ has a decomposition
\[
\A(I) =\int_{X}^{\oplus}\A_{\lambda}(I)\text{d}\mu(\lambda)
\]
where the nets $\A_{\lambda}$ are M\"{o}bius (resp. diffeomorphism) 
covariant and irreducible. The decomposition is 
unique (up to a set of measure 0). Here we have set 
$Z=L^{\infty}(X,\mu)$\footnote{If $\H$ is non separable the 
decomposition should be stated in a more general form.}.
\end{lemma}

\begin{proof}
Assume $\A$ to  be M\"{o}bius covariant.
Given a vector $\xi\in\H$, $U(\Lambda_I (t))\xi=\xi, \ \forall t\in\mathbb 
R$, iff $U(g)\xi=\xi, \ \forall g\in\PSL$, see
\cite{GL2}. Hence if $I\subset\tilde I$ are intervals and 
$A\in\A(\tilde I)$, the vector $A\Om$ is fixed by $U(\Lambda_I 
(\cdot))$ iff it is fixed by $U(\Lambda_{\tilde I} 
(\cdot))$. Thus $A$ is fixed by the modular group of $(\A(I),\Om)$ iff 
it is fixed by the modular group of $(\A(\tilde I),\Om)$. In other 
words the centralizer $Z_{\omega}$ of $\A(I)$ is independent of $I$ 
hence, by locality, it is contained in the center of any $\A(I)$. 
Since the center is always contained in the centralizer, it follows 
that $Z_{\omega}$ must be the common center of all the $\A(I)$'s. 
The statement is now an immediate consequence of the uniqueness of the 
direct integral decomposition of a von Neumann algebra into factors.

If $\A$ is further diffeomorphism covariant, then the fiber 
$\A_{\lambda}$ in the decomposition is diffeomorphism covariant 
too. Indeed $\Diff(I)\subset\A(I)$ decomposes through the space $X$ and so 
does $\Diff(S^1)$, which
is generated by $\{\Diff(I), I\in\I\}$ (cf. e.g. \cite{Lk}). 
\end{proof}

Before concluding this subsection, we explicitly say that 
two conformal nets $\A_1$ and $\A_2$ are  {\it isomorphic} if there is 
a unitary $V$ from the Hilbert space of $\A_1$ to the Hilbert space of
$\A_2$, mapping the vacuum vector of $\A_1$ to the vacuum 
vector of $\A_2$, such that  $V\A_1(I)V^*=\A_2(I)$ for all $I\in\I$. 
Then $V$ also intertwines the M\"{o}bius
covariance representations of $\A_1$ and $\A_2$ \cite{BGL},
because of the uniqueness of these representations due to eq. (\ref{BW}). 
Our classification will be up to isomorphism. Yet, as a consequence of our 
results, our classification will indeed be up to the 
priori weaker notion of isomorphism where $V$ is not assumed to 
preserve the vacuum vector.
 
Note also that, by Haag duality, two fields 
generate isomorphic nets iff they are relatively local, namely
belong to the same Borchers class (see \cite{H}).

\subsubsection{Representations}
\label{Rep}

Let $\A$ be an irreducible local M\"{o}bius covariant (resp.
conformal) net. A {\it representation} 
$\pi$ of $\A$ is a map
\[
I\in\I\to\pi_I\ ,
\]
where $\pi_I$ is a representation of $\A(I)$ on a fixed Hilbert space 
$\H_{\pi}$ such that
\[ 
\pi_{\tilde I}\res_{\A(I)}=\pi_I, \quad I\subset\tilde I\ .
\]
We shall always implicitly assume that $\pi$ is locally normal, namely
$\pi_I$ is normal for all $I\in\I$, which is automatic if $\H_{\pi}$ is 
separable \cite{T}.

We shall say that $\pi$ is  M\"{o}bius (resp. conformal) covariant
if there exists a positive energy representation $U_{\pi}$ of 
$\PSL^{\tilde{}}$ (resp. of $\Diff(S^1)$) such that
\[
U_{\pi}(g)\A(I)U_{\pi}(g)^{-1}= \A(gI),
\ g \in\PSL^{\tilde{}}\quad (\text{resp.}\ 
g\in\Diff(S^1) ).
\]
(Here $\PSL^{\tilde{}}$ denotes the universal central cover of $\PSL$.)
The identity representation of $\A$ is called the {\it vacuum} 
representation; if convenient, it will be denoted by $\pi_0$.

We shall say that a representation $\r$ is {\it localized} in a 
interval $I_0$ if $\H_{\r}=\H$ and $\rho_{I'_0}={\rm id}$. Given an 
interval $I_0$ and a representation $\pi$ on a separable Hilbert 
space, there is a representation $\rho$ unitarily equivalent to $\pi$ 
and localized in $I_0$. This is due the type III factor property.
If $\r$ is a representation localized in $I_0$, then by Haag duality 
$\r_I$ is an endomorphism of $\A(I)$ if $I\supset I_0$. 
The endomorphism $\r$ is called a 
DHR endomorphism \cite{DHR} localized in $I_0$. The {\it index} of a 
representation $\r$ is the Jones index $[\r_{I'}(\A(I'))':\r_I(\A(I))]$ 
for any interval $I$ or, equivalently,
the Jones index $[\A(I):\r_I(\A(I))]$ of $\r_I$, if $I\supset I_0$. 
The (statistical) {\it dimension} $d(\rho)$ of $\r$ is the square 
root of the index.

The unitary equivalence $[\r]$ class of a representation $\r$ of $\A$ 
is called a {\it sector} of $\A$.

\subsubsection{Subnets}

Let $\A$ be a M\"{o}bius covariant (resp. conformal) net on $S^1$ and 
$U$ the unitary covariance representation of the M\"{o}bius group 
(resp. of $\Diff(S^1)$). 

A M\"{o}bius covariant (resp. conformal) {\it subnet} $\B$ of $\A$ is 
an isotonic map $I\in\I\to\B(I)$ that associates to each interval $I$ a von 
Neumann subalgebra $\B(I)$ of $\A(I)$ with $U(g)\B(I)U(g)^*=\B(gI)$ 
for all $g$ in the M\"{o}bius group (resp. in $\Diff(S^1)$). 

If $\A$ is local and irreducible, then the modular group of 
$(\A(I),\Om)$ is ergodic and so is its restriction to $\B(I)$, thus 
the each $\B(I)$ is a factor. By the Reeh-Schlieder theorem the 
Hilbert space $\H_0\equiv\overline{\B(I)\Om}$ is independent of $I$. 
The restriction of $\B$ to $\H_0$ is then an irreducible local M\"{o}bius 
covariant (resp. conformal) net on $\H_0$ and we denote it here by 
$\B_0$. The vector $\Om$ is separating for $\B(I)$ therefore the map 
$B\in\B(I)\to B|_{\H_0}\in\B_0(I)$ is an isomorphism. Its inverse 
thus defines a representation of $\B_0$, that we shall call the restriction 
to $\B$ of the vacuum representation of $\A$ (as a sector this is 
given by the dual canonical 
endomorphism of $\A$ in $\B$). Indeed we shall sometimes 
identify $\B(I)$ and $\B_0(I)$ although, properly speaking, $\B$ is 
not a M\"{o}bius covariant net because $\Om$ is not cyclic.

If $\B$ is a subnet of $\A$ we shall denote here $\B''$ the von Neumann algebra 
generated by all the algebras $\B(I)$ as $I$ varies in the intervals $\I$. 
The subnet $\B$ of $\A$ is said to be {\it irreducible} if 
$\B'\cap\A(I)=\mathbb C$ (if $\B$ is strongly additive this is 
equivalent to $\B(I)'\cap\A(I)=\mathbb C$). Note that an irreducible 
subnet is not an irreducible net. If $[\A:\B]<\infty$ then $\B$ is 
automatically irreducible.

The following lemma will be used in the paper. 
\begin{lemma}\label{subnet}
Let $\A$ be a M\"{o}bius covariant net on $S^1$ and $\B$ a M\"{o}bius covariant 
subnet. Then $\B''\cap\A(I)=\B(I)$ for any given $I\in\I$. 
\end{lemma}
\begin{proof}
By eq. (\ref{BW}) $\B(I)$ is globally 
invariant under the modular group of $(\A(I),\Om)$, thus by Takesaki's theorem 
there exists a vacuum preserving conditional expectation from $\A(I)$ 
to $\B(I)$ and an operator $A\in\A(I)$ belongs to $\B(I)$ if and 
only if $A\Om\in\overline{\B(I)\Om}$. By the Reeh-Schlieder theorem 
$\overline{\B''\Om}=\overline{\B(I)\Om}$ and this immediately entails 
the statement.
\end{proof}

\subsection{Virasoro algebras and Virasoro nets}
\label{Vir-alg-net}

The Virasoro algebra is the infinite dimensional Lie algebra generated 
by elements $\{L_n \mid n\in\mathbb Z\}$ and $c$ with
relations
\begin{equation}\label{vir-rel}
[L_m,L_n]=(m-n) L_{m+n} + \frac{c}{12}(m^3-m)\delta_{m,-n}.
\end{equation}
and $[L_n,c]=0$. It is the (complexification of) the unique, non-trivial
one-dimensional central extension of the Lie algebra of $\Diff(S^1)$.

We shall only consider unitary 
representations of the Virasoro algebra 
(i.e. $L_n^*=L_{-n}$ in the representation space) with positive energy
 (i.e. $L_{0}>0$ in the representation space)
indeed the ones associated with a projective unitary representation of $\Diff(S^1)$.

In any irreducible representation the {\it central charge} $c$ is 
a scalar, indeed $c=1-6/m(m+1)$, $(m=2,3,4,\dots)$ or $c\geq 1$
\cite{FQS} and all these values are allowed \cite{GKO}.

For every admissible value of $c$ there is exactly one irreducible 
(unitary, positive energy) representation $U$ of the Virasoro 
algebra (i.e. projective unitary representation  
of $\Diff(S^1)$) such that the lowest eigenvalue of the conformal 
Hamiltonian $L_0$ (i.e. the {\it spin}) is 0; this is the vacuum 
representation with central charge 
$c$. One can then define the Virasoro net
\[
\Vir_c(I)\equiv U(\Diff(I))''.
\]
Any other projective unitary
irreducible representation of $\Diff(S^1)$ with a given central 
charge $c$ is uniquely determined 
by its spin. Indeed, as we shall see, 
these representations with central charge $c$
correspond bijectively to the irreducible representations (in the 
sense of Subsection \ref{Rep}) of
the $\Vir_c$ net, namely their equivalence classes correspond
to the irreducible sectors of the $\Vir_c$ net.

In conformal field theory, the $\Vir_c$ net for $c<1$ are studied 
under the name of {\it minimal models} (see \cite[Chapters 7--8]{DMS}, for
example). Notice that they are indeed minimal in the sense they 
contain no non-trivial subnet \cite{C}.

For the central charge $c=1-6/m(m+1)$, $(m=2,3,4,\dots)$,
we have $m(m-1)/2$  characters $\chi_{(p,q)}$ of the
minimal model labeled with
$(p,q)$, $1\le p\le m-1$, $1\le q\le m$ with the identification
$\chi_{(p,q)}=\chi_{(m-p,m+1-q)}$, as in \cite[Subsection 7.3.4]{DMS}.
They have fusion rules as in \cite[Subsection 7.3.3]{DMS} and
they are given as follows.
\begin{equation}\label{fusion}
\chi_{(p,q)}\chi_{(p',q')}=
\bigoplus_{r=|p-p'|+1, r+p+p':{\rm odd}}
^{\min(p+p'-1, 2m-p-p'-1)\hphantom{x}}
\bigoplus_{s=|q-q'|+1, s+q+q':{\rm odd}}
^{\hphantom{x}\min(q+q'-1, 2(m+1)-q-q'-1)}
\chi_{(r,s)}
\end{equation}
Note that here the product $\chi_{(p,q)}\chi_{(p',q')}$ denotes the 
fusion of characters and not their pointwise product as functions.

For the character $\chi_{(p,q)}$, we have a spin
\begin{equation}
\label{spin}
h_{p,q}=\frac{((m+1)p-mq)^2-1}{4m(m+1)}
\end{equation}
by \cite{GKO}. (Also see \cite[Subsection 7.3.3]{DMS}.)
The characters $\{\chi_{(p,q)}\}_{p,q}$ have the $S$, $T$-matrices
of Kac-Petersen as in \cite[Section 10.6]{DMS}.

\subsection{Virasoro nets and classification of the modular invariants}

Cappelli-Itzykson-Zuber \cite{CIZ} and Kato \cite{Kt} have made
an $A$-$D$-$E$ classification of the modular invariant matrices
for $SU(2)_k$.  That is, for the unitary representation of the
group $SL(2,\Z)$ arising from $SU(2)_k$ as in
\cite[Subsection 17.1.1]{DMS}, they classified matrices $Z$
with non-negative integer entries in the commutant of this unitary
representations, up to the normalization $Z_{00}=1$.   Such
matrices are called {\sl modular invariants} of $SU(2)_k$ and
labeled with Dynkin diagrams $A_n$, $D_n$,
$E_{6,7,8}$ by looking at the diagonal entries of the matrices
as in the table (17.114) in \cite{DMS}.
Based on this classification, Cappelli-Itzykson-Zuber
\cite{CIZ} also gave a classification of the modular invariant
matrices for the above minimal models and the unitary representations
of $SL(2,\Z)$ arising from the $S$, $T$-matrices mentioned at the
end of the previous subsection.  From 
our viewpoint, we will regard this as a classification of matrices
with non-negative integer entries in the commutant of the unitary
representations of $SL(2,\Z)$ arising from the Virasoro net
$\Vir_c$ with $c<1$.  Such modular invariants of the minimal models
are labeled with pairs of Dynkin diagrams of $A$-$D$-$E$ type
such that the difference of their Coxeter numbers is 1.  The
classification tables are given in Table \ref{Vir-mod-I} for
so-called type I (block-diagonal) modular invariants, where each modular
invariant  $(Z_{(p,q),(p',q')})_{(p,q),(p',q')}$ is listed
in the form $\sum Z_{(p,q),(p',q')}
\chi_{(p,q)} \overline{\chi_{(p',q')}}$,
and we refer to
\cite[Table 10.4]{DMS} for the type II modular invariants, since
we are mainly concerned with type I modular invariants in this
paper.  (Note that the coefficient $1/2$ in the table arises from
a double counting due to the identification $\chi_{(p,q)}=
\chi_{(m-p,m+1-q)}$.)  Here the labels come from the diagonal entries
of the matrices again, but we will give our subfactor interpretation
of this labeling later.

\begin{table}[htbp]
\begin{center}{\footnotesize
\begin{tabular}{|c|c|}\hline
Label & $\displaystyle\sum_{\vphantom{x}}^{\vphantom{x}}
Z_{(p,q),(p',q')} \chi_{(p,q)} \overline{\chi_{(p',q')}}$
\\ \hline
$(A_{n-1},A_n)$ & $\displaystyle\sum_{p,q}^{\vphantom{x}}
|\chi_{(p,q)}|^2/2$
\\ \hline
$(A_{4n},D_{2n+2})$
& $\displaystyle\sum_{q:{\rm\ odd}}^{\vphantom{x}}
|\chi_{(p,q)}+\chi_{(p,4n+2-q)}|^2/2$
\\ \hline
$(D_{2n+2},A_{4n+2})$ &
$\displaystyle\sum_{p:{\rm\ odd}}^{\vphantom{x}}
|\chi_{(p,q)}+\chi_{(4n+2-p,q)}|^2/2$
\\ \hline
$(A_{10},E_6)$ & 
$\displaystyle\sum_{p=1}^{10} \left\{|\chi_{(p,1)}+\chi_{(p,7)}|^2+
|\chi_{(p,4)}+\chi_{(p,8)}|^2+|\chi_{(p,5)}+\chi_{(p,11)}|^2\right\}/2$
\\ \hline
$(E_6, A_{12})$ &
$\displaystyle\sum_{q=1}^{12} \left\{|\chi_{(1,q)}+\chi_{(7,q)}|^2+
|\chi_{(4,q)}+\chi_{(8,q)}|^2+|\chi_{(5,q)}+\chi_{(11,q)}|^2\right\}/2$
\\ \hline
$(A_{28},E_8)$ &
$\displaystyle\sum_{p=1}^{28} \left\{|\chi_{(p,1)}+\chi_{(p,11)}+
\chi_{(p,19)}+\chi_{(p,29)}|^2+|\chi_{(p,7)}+\chi_{(p,13)}+
\chi_{(p,17)}+\chi_{(p,23)}|^2\right\}/2$
\\ \hline
$(E_8, A_{30})$ &
$\displaystyle\sum_{q=1}^{30} \left\{|\chi_{(1,q)}+\chi_{(11,q)}+
\chi_{(19,q)}+\chi_{(29,q)}|^2+|\chi_{(7,q)}+\chi_{(13,q)}+
\chi_{(17,q)}+\chi_{(23,q)}|^2\right\}/2$
\\ \hline
\end{tabular}
\caption{Type I modular invariants of the minimal models}
\label{Vir-mod-I}
}\end{center}
\end{table}

\subsection{$Q$-systems and classification}
\label{Q-syst}

Let $M$ be an infinite factor. 
A $Q$-system $(\rho, V, W)$ in \cite{L2} is a triple of an
endomorphism of $M$ and isometries $V\in\Hom(\id,\rho)$,
$W\in\Hom(\rho,\rho^2)$ satisfying the following identities:
\begin{eqnarray*}
&&V^* W =\rho(V^*)W \in\R_+,\\
&&\rho(W)W=W^2.
\end{eqnarray*}
The abstract notion of $Q$-system for tensor categories is contained 
in \cite{LRo}. (We had another identity in addition to the above
in \cite{L2} as the definition of a
$Q$-system, but it was proved to be redundant in \cite{LRo}.)

If $N\subset M$ is a finite-index subfactor, the 
associated canonical endomorphism gives rise to a $Q$-system. 
Conversely any $Q$-system determines a subfactor $N$ of $M$ such that
$\rho$ is the canonical endomorphism for $N\subset M$: $N$ is given by
\[
N=\{x\in M\mid Wx=\rho(x)W\}.
\]

We say $(\rho, V, W)$ is irreducible when $\dim\Hom(\id,\rho)=1$.
We say that two $Q$-systems $(\rho, V_1, W_1)$ and $(\rho, V_2, W_2)$
are {\sl equivalent} if we have a unitary $u\in \Hom(\rho,\rho)$ satisfying
\[
V_2=u V_1,\qquad W_2=u\rho(u)W_1 u^*.
\]
This equivalence of $Q$-systems is equivalent to inner conjugacy of
the corresponding subfactors.

Subfactors $N\subset M$ and extensions $\tilde M\supset M$ of $M$ are 
naturally related by Jones basic construction (or by the canonical 
endomorphism).
The problem we are interested in is a classification of $Q$-systems
up to equivalence when a system of endomorphisms is given
and $\rho$ is a direct sum of endomorphisms in the system.

\subsection{Classification of local extensions of the $SU(2)_k$ net}

As a preliminary to our main classification theorem, we first
deal with local extensions of the $SU(2)_k$ net.
The $SU(n)_k$ net was constructed in \cite{W} using a representation
of the loop group \cite{PS}.  By the results on the fusion rules
in \cite{W} and the spin-statistics theorem \cite{GL2},
we know that the usual $S$- and $T$-matrices of $SU(n)_k$ as 
in \cite[Section 17.1.1]{DMS} and those arising from the braiding
on the $SU(n)_k$ net as in \cite{R1} coincide.

We start with the following result.

\begin{proposition}
\label{finite-index}
Let $\N$ be a M\"{o}bius covariant net on the circle. Suppose that
$\N$ admits only finitely many irreducible DHR sectors and each sector 
is sum of sectors with finite statistical dimension.  If
$\M$ is an irreducible local extension of $\N$, then 
the index $[\M:\N]$ is finite.
\end{proposition}

\begin{proof}
As in \cite[Lemma 13]{L4}, we have a vacuum preserving
conditional expectation $\M(I)\to \N(I)$.  The dual canonical endomorphism
$\th$ for $\N(I)\subset \M(I)$ decomposes into DHR endomorphisms of the
net $\N$, but we have only finitely many such endomorphisms of
finite statistical dimensions by assumption.
Then the result in \cite[page 39]{ILP} shows that multiplicity of
each such DHR endomorphism in $\th$ is finite, thus the index 
($=d(\th)$) is also finite.
\end{proof}

We are interested in the classification problem of irreducible local
extensions $\M$ when $\N$ is given.  (Note that if 
we have finite index $[\M:\N]$, then the irreducibility holds
automatically by \cite[I, Corollary 3.6]{BE}, \cite{DRL}.)
The basic case of this problem is the one where
$\N(I)$ is given from $SU(2)_k$ as in \cite{W}.
In this case, the following classification result is implicit
in \cite{BEK2}, but for the sake of completeness, we state and give
a proof to it here as follows.  Note that $G_2$ in Table \ref{SU2-ext}
means the exceptional Lie group $G_2$.

\begin{theorem}
\label{class-SU2}
The irreducible local extensions of the $SU(2)_k$ net are in
a bijective correspondence to the Dynkin diagrams of
type $A_n$, $D_{2n}$, $E_6$, $E_8$ as in Table \ref{SU2-ext}.
\end{theorem}

\begin{table}[htbp]
\begin{center}
\begin{tabular}{|c|c|c|}\hline
level $k$ & Dynkin diagram & Description \\ \hline
$n-1, (n\ge1)$ & $A_n$ & $SU(2)_k$ itself \\ \hline
$4n-4, (n\ge2)$ & $D_{2n}$ & Simple current extension of index 2 \\ \hline
10 & $E_6$ &  Conformal inclusion $SU(2)_{10}\subset SO(5)_1$ \\ \hline
28 & $E_8$ &  Conformal inclusion $SU(2)_{28}\subset (G_2)_1$ \\ \hline
\end{tabular}
\caption{Local extensions of the $SU(2)_k$ net}
\label{SU2-ext}
\end{center}
\end{table}

\begin{proof}
The $SU(2)_k$ net $\N$ is
completely rational by \cite{X3}, thus any local
extension $\M$ is of finite index by \cite[Corollary 39]{KLM}
and Proposition \ref{finite-index}.
For a fixed interval $I$, we have a subfactor $\N(I)\subset \M(I)$ and
can apply the $\a$-induction for the system $\Delta$ of DHR endomorphisms
of $\N$.  Then the matrix $Z$ given by
$Z_{\la\,\mu}=\lan \a^+_\la,\a^-_\mu\ran$
is a modular invariant for $SU(2)_k$
by \cite[Corollary 5.8]{BEK1} and thus one of
the matrices listed in \cite{CIZ}.  Now we have locality of $\M$, so 
we have $Z_{\la,0}=\lan \a^+_\la,\id\ran=\lan \la,\th\ran$,
where $\th$ is the dual canonical endomorphism for $\N(I)\subset \M(I)$
by \cite{X1}, and the modular invariant matrix $Z$ must be
block-diagonal, which is said to be of type I as in Table \ref{Vir-mod-I}.
Looking at the classification of \cite{CIZ}, we have only the following
possibilities for $\th$.
\begin{eqnarray*}
\th&=&\id,\quad
\text{for the type $A_{k+1}$ modular invariant at level $k$},\\
\th&=&\la_0\oplus\la_{4n-4},\quad
\text{for the type $D_{2n}$ modular invariant at level $k=4n-4$},\\
\th&=&\la_0\oplus\la_6,\quad
\text{for the type $E_6$ modular invariant at level $k=12$},\\
\th&=&\la_0\oplus\la_{10}\oplus\la_{18}\oplus\la_{28},\quad
\text{for the type  $E_8$ modular invariant at level $k=28$}.
\end{eqnarray*}

By \cite{X1}, \cite[II, Section 3]{BE}, we know that all these cases
indeed occur, and we have the unique $Q$-system for each case by 
\cite[Section 6]{KO}.
(In \cite[Definition 1.1]{KO}, Conditions 1 and 3 correspond to the
axioms of the $Q$-system in Subsection \ref{Q-syst}, Condition 4
corresponds to irreducibility, and Condition 3 corresponds to
chiral locality in \cite[Theorem 4.9]{LR}
in the sense of \cite[page 454]{BEK1}.)
By \cite[Theorem 4.9]{LR}, we conclude that the local
extensions are classified as desired.
\end{proof}

\begin{remark}{\rm
The proof of uniqueness for the $E_8$ case in \cite[Section 6]{KO}
uses vertex operator algebras.  Izumi has recently given a direct
proof of uniquenss of the $Q$-system using an intermediate
extension.  We have later further obtained another proof based
on 2-cohomology vanishing for the tensor category $SU(2)_k$ in \cite{KL}.
An outline of the arguments is as follows.

Suppose we have two $Q$-systems for this dual canonical endomorphism
of an injective type III$_1$ factor $M$.  We need to prove that
the two corresponding subfactors $N_1\subset M$ and $N_2\subset M$
are inner conjugate.  First, it is easy to prove that the paragroups
of these two subfactors are isomorphic to that of the Goodman-de
la Harpe-Jones subfactor \cite[Section 4.5]{GHJ} arising from $E_8$.
Thus we may assume that these two subfactors are conjugate.  From
this, one shows that
the two $Q$-systems differ only by a ``2-cocycle'' of the even
part of the tensor category $SU(2)_{28}$.  Using the facts that
the fusion rules of $SU(2)_k$ have no multiplicities and that
all the $6j$-symbols are non-zero, one proves that any such 2-cocycle
is trivial.  This implies that the two $Q$-systems are equivalent.
}\end{remark}

\section{The Virasoro nets as cosets}
\label{Virasoro=coset}

Based on the coset construction of projective unitary representations of
the Virasoro algebras with central charge less than 1 by
Goddard-Kent-Olive \cite{GKO}, it is natural to expect that
the Virasoro net on the circle with central charge
$c=1-6/m(m+1)$ and the coset model arising from the diagonal
embedding $SU(2)_{m-1}\subset SU(2)_{m-2}\times SU(2)_1$
as in \cite{X4} are isomorphic.  We prove this isomorphism
in this section.  This, in particular, implies that the
Virasoro nets with central charge less than 1 are completely
rational in the sense of \cite{KLM}.

\begin{lemma}\label{diffeo1}
If $\A$ is a $\Vir$ net, then every M\"{o}bius covariant
representation $\pi$ of $\A$ is $\Diff(S^1)$ covariant.
\end{lemma}

\begin{proof}
Indeed $\A(I)$ is generated by
$U(\Diff(I))$, where $U$ is an irreducible projective unitary representation 
of $\Diff(S^1)$, and $U(g)$ clearly implements the covariance 
action of $g$ on $\A$ if $g$ belongs to $\Diff(I)$. 
Thus $\pi_I(U(g))$ implements the covariance action of $g$ in the 
representation $\pi$. 
As $\Diff(S^1)$ is generated by $\Diff(I)$ as $I$
varies in the intervals, the full $\Diff(S^1)$ acts covariantly.
The positivity of the energy holds by the M\"{o}bius covariance 
assumption.
\end{proof}

\begin{lemma}
\label{lemma-expans}
Let $\A$ be an irreducible M\"{o}bius covariant local net, $\B$ and 
$\CC$ mutually commuting subnets of $\A$. Suppose the restriction to 
$\B\vee\CC\simeq\B\otimes\CC$ of the vacuum representation $\pi_0$ of $\A$ 
has the (finite or infinite) expansion
\begin{equation}\label{expans}
\pi_0 |_{\B\vee\CC} =\bigoplus^n_{i=0} \rho_i\otimes\s_i ,
\end{equation}
where $\rho_0$ is the vacuum representation of $\B$,
$\s_0$ is the vacuum representation of $\CC$, and
$\rho_0$ is disjoint from $\rho_i$ if $i\neq 0$. 
Then $\CC(I)=\B'\cap\A(I)$.
\end{lemma}

\begin{proof}
The Hilbert space $\H$ of $\A$ decomposes according to the expansion 
(\ref{expans}) as
\[
\H =\bigoplus_{i=0}^n \H_i\otimes\K_i.
\]
The vacuum vector $\Om$ of $\A$ corresponds to 
$\Om_\B\otimes\Om_\CC\in\H_0\otimes\K_0$, where $\Om_\B$ and $\Om_\CC$ 
are the vacuum vector of $\B$ and $\CC$, because $\H_0\otimes\K_0$ 
is, by assumption, the support of the representation $\rho_0\otimes\s_0$.
We then have
\[
\pi_0(B) =\sum_{i=0}^n \rho_i(B)\otimes 1|_{\K_i},\quad B\in \B(I).
\]
and, as $\rho_0$ is disjoint from $\rho_i$ if $i\neq 0$, 
\[
\pi_0(\B)' = (1_{\H_0}\otimes B(\K_0)) \oplus \cdots
\]
where we have set $\pi_0(\B)'\equiv(\bigvee_{I\in\I}\B(I))'$ and the 
dots stay for operators on the orthogonal complement of 
$\H_0\otimes\K_0$. 
It follows that if $X\in\pi_0(\B)'$, then $X\Om\in\H_0\otimes\K_0$.

With $\L$ the subnet of $\A$ given by $\L(I)\equiv\B(I)\vee\CC(I)$, 
we then have by the Reeh-Schlieder theorem
\[
X\in\pi_0(\B)'\cap\A(I)\implies X\Om\in 
\overline{\L(I)\Om}\implies X\in \L(I),
\]
where the last implication follows by Lemma \ref{subnet}.
As $\L(I)\simeq \B(I)\otimes\CC(I)$ and $X$ commutes with $\B(I)$, we 
have $X\in\CC(I)$ as desired.
\end{proof} 

The proof of the following corollary has been indicated to the 
authors (independently) by F. Xu and and S. Carpi. 
Concerning our original proof, see Remark \ref{opr}
at the end of this section.

\begin{corollary}
\label{Vir-coset}
The Virasoro net on the circle with central charge
$c=1-6/m(m+1)$ and the coset net arising from the diagonal
embedding $SU(2)_{m-1}\subset SU(2)_{m-2}\times SU(2)_1$
are isomorphic.
\end{corollary}

\begin{proof}
As shown in \cite{GKO}, $\Vir_c$ is a subnet of the above coset net for
$c=1-6/m(m+1)$. Moreover formula in  \cite[(2.20)]{GKO},
obtained by comparison of 
characters, shows in particular that the hypothesis in Lemma 
\ref{lemma-expans} hold true with $\A$ the $SU(2)_{m-2}\times SU(2)_1$ net,
$\B$ the $SU(2)_{m-1}$ subnet (coming from diagonal embedding) and 
$\CC$ the $\Vir_c$ subnet. Thus the corollary follows.
\end{proof}

\begin{corollary}
\label{coset-rat}
The Virasoro net on the circle $\Vir_c$ with central charge $c<1$
is completely rational.
\end{corollary}

\begin{proof}
The Virasoro net on the circle $\Vir_c$ with central charge $c=1-6/m(m+1)$
coincides with the coset net arising from the diagonal
embedding $SU(2)_{m-1}\subset SU(2)_{m-2}\times SU(2)_1$ by
Corollary \ref{Vir-coset}, thus it is completely rational by
\cite[Sect. 3.5.1]{L4}.
\end{proof}
Next proposition shows in particular that the central charge is 
defined for any local irreducible conformal net.

\begin{proposition}
\label{Vir-sub}
Let $\M$ be a local irreducible conformal net on the circle.
Then it contains canonically a Virasoro net as a subnet.  If its central
charge $c$ satisfies $c<1$, then the Virasoro subnet is an irreducible 
subnet with finite index.
\end{proposition}

\begin{proof}
Let $U$ be the projective unitary representation of $\Diff(S^1)$ implementing the 
diffeomorphism covariance on $\M$ and set
\[
\M_{\Vir}(I)=U(\Diff(I))''.
\]
Then $U$ is the direct sum the vacuum representation of $\Vir_c$ and 
another representation of $\Vir_c$. Indeed, as $\M_{\Vir}$ is a subnet 
of $\M$, all the subrepresentation of $\M_{\Vir}$ are mutually locally 
normal, so they have the same central charge $c$.
Note that the central charge is well defined because $U$ is a 
projective unirary representation.

Suppose now that $c<1$.
For an interval $I$ we must show that $\M_{\Vir}(I)'\cap \M(I) = \C$.
By locality it is enough to show that
$(\M_{\Vir}(I')\vee  \M_{\Vir}(I))'\cap \M(I) = \C$.
Because the net $\Vir$ is completely rational by
Corollary \ref{coset-rat}, it is strongly additive
in particular, and thus we have
$\M_{\Vir}(I')\vee  \M_{\Vir}(I)$
is equal to the weak closure of all the net $\M_{\Vir}$.
Then any $X$ in $\M(I)$ that
commutes with $\M_{\Vir}(I')\vee\M_{\Vir}(I)$ would commute 
with $U(g)$ for any $g$ in
$\Diff(I)$ for every interval $I$. Now the group $\Diff(S^1)$
is generated by  the subgroups $\Diff(I)$, so $X$ would commute with all
$U(\Diff(S^1) )$, in particular it would be fixed 
by the modular group of $(\M(I),\Om)$,
which is ergodic, thus $X$ is to be a scalar.

Then $[\M:\M_{\Vir}]<\infty$ by Prop. \ref {finite-index} 
and Corollary \ref{coset-rat}.
\end{proof}

We remark that we can also prove
that $\M_{\Vir}(I')\vee \M_{\Vir}(I)$ and the range of
full net $\M_{\Vir}$ have the same weak closure as follows.
Since $\M_{\Vir}$ is obtained as a direct 
sum of irreducible sectors
$\rho_i$ of $\M_{\Vir}$ localizable in $I$, it is enough to show 
that the intertwiners between
$\rho_i$ and $\rho_j$ as endomorphisms of the factor $\Vir_c(I)$
are the same as the
intertwiners between $\rho_i$ and $\rho_j$
as representations of $\Vir_c$. Since each $\rho_i$
has a finite index by complete rationality as in \cite[Corollary 39]{KLM},
the result follows by the theorem of
equivalence of local and global intertwiners in \cite{GL2}.

Given a local irreducible conformal net $\M$, the subnet $\M_{\Vir}$ 
constructed in Proposition \ref{Vir-sub} is the {\it Virasoro subnet} 
of $\M$. It is isomorphic to $\Vir_c$ for some $c$, except that the 
vacuum vector is not cyclic. Of course, if $\M$ is a Virasoro net, 
then $\M_{\Vir}=\M$ by construction.

Xu has constructed irreducible DHR endomorphisms of the coset net arising
from the diagonal embedding $SU(n)\subset SU(n)_k\otimes SU(n)_l$
and computed their fusion rules in \cite[Theorem 4.6]{X4}.  In the
case of the Virasoro net with central charge $c=1-6/m(m+1)$,
this gives the following result.  For 
$SU(2)_{m-1}\subset SU(2)_{m-2}\times SU(2)_1$, we use a label
$j=0,1,\dots,m-2$ for the irreducible DHR endomorphisms of $SU(2)_{m-2}$.
Similarly, we use $k=0,1,\dots,m-1$ and $l=0,1$ for the
irreducible DHR endomorphisms of $SU(2)_{m-1}$ and $SU(2)_1$,
respectively.  (The label ``0'' always denote the identity endomorphism.)
Then the irreducible DHR endomorphisms of the Virasoro net are
labeled with triples $(j,k,l)$ with $j-k+l$ being even under
identification $(j,k,l)=(m-2-j,m-1-k,1-l)$.  
Since $l\in\{0,1\}$ is uniquely determined by $(j,k)$ under this
parity condition, we may and do label them with pairs $(j,k)$ under
identification $(j,k)=(m-2-j,m-1-k)$.  In order to identify
these DHR endomorphisms with characters of the minimal models,
we use variables $p,q$ with $p=j+1, q=k+1$.  Then we have
$p\in\{1,2,\dots,m-1\}$, $q\in\{1,2,\dots,m\}$.  We denote the
DHR endomorphism of the Virasoro net labeled with the pair $(p,q)$ 
by $\la_{(p,q)}$.  That is, we have $m(m-1)/2$ irreducible DHR sectors
$[\la_{(p,q)}]$, $1\le p\le m-1$, $1\le q\le m$ with the identification
$[\la_{(p,q)}]=[\la_{(m-p,m+1-q)}]$, and then their fusion rules are
identical to the one in $(\ref{fusion})$.
Although the indices of these DHR sectors are not explicitly computed
in \cite{X4}, these fusion rules uniquely determine the indices by the
Perron-Frobenious theorem.
All the irreducible DHR sectors of the Virasoro net on the circle with
central charge $c=1-6/m(m+1)$ are given as $[\la_{(p,q)}]$ as above
by \cite[Proposition 3.7]{X5}.  Note that the $\mu$-index of 
the Virasoro net with central charge $c=1-6/m(m+1)$ is
$$\frac{m(m+1)}{8\sin^2{\frac{\pi}{m}} \sin^2{\frac{\pi}{m+1}}}$$
by \cite[Lemma 3.6]{X5}.

Next we need statistical phases of the DHR sectors $[\la_{(p,q)}]$.
Recall that an irreducible DHR endomorphism $r\in\{0,1,\dots,n\}$
of $SU(2)_n$ has the statistical phase
$\exp (2\pi r(r+2)i/4(n+2))$.  This shows that for the triple
$(j,k,l)$, the statistical phase of the DHR endomorphism $l$ of
$SU(2)_1$ is given by $\exp (2\pi (j-k)^2 i/4)$, because of the
condition $j-k+l\in 2\Z$.  Then by \cite[Theorem 4.6.(i)]{X6}
and \cite[Lemma 6.1]{BE4}, we obtain that
the statistical phase of the DHR endomorphism $[\la_{(p,q)}]$ is 
$$\exp 2\pi i \left(\dfrac{(m+1)p^2-mq^2-1+m(m+1)(p-q)^2}{4m(m+1)}\right),$$
which is equal to $\exp(2\pi ih_{p,q})$
with $h_{p,q}$ as in (\ref{spin}).
Thus the $S$, $T$-matrices of Kac-Petersen
in \cite[Section 10.6]{DMS} and the $S$, $T$-matrices for the DHR sectors
$[\la_{(p,q)}]$ defined from the braiding as in \cite{R1} coincide.
This shows that the unitary representations of $SL(2,\Z)$ 
studied in \cite{CIZ} for the minimal models and those arising 
from the braidings on the Virasoro nets are identical.  
So when we say the modular invariants for the
Virasoro nets, we mean those in \cite{CIZ}.

\begin{corollary}
\label{net-diff}
There is a natural bijection between 
representations of the $\Vir_c$ 
net and projective unitary (positive energy) representations of the group 
$\Diff(S^1)$ with central charge $c<1$.
\end{corollary}

\begin{proof}
If $\pi$ is a representation of $\Vir_c$, then the irreducible sectors 
are automatically M\"{o}bius covariant with 
positivity of the energy \cite{GL1} 
because the they have finite index and $\Vir_c$ is strongly additive
by Cor. \ref{coset-rat}. Thus all sectors are diffeomorphism 
covariant by Lemma \ref{diffeo1} and the associated covariance 
representation $U_{\pi}$ is a projective unitary representation of $\Diff(S^1)$.
The converse follows from the above description of the DHR sectors.
\end{proof}

\begin{remark}\label{opr}{\rm
We give a remark about the thesis \cite{Lk}
of Loke.  He constructed irreducible DHR endomorphisms of the Virasoro
net with $c<1$ using the discrete series of projective unitary 
representations of 
$\Diff(S^1)$ and computed their fusion rules, which coincides with the one
given above.  However, his proof of strong additivity contains a serious
gap and this affects the entire results in \cite{Lk}.
So we have avoided using his results here. (The proof of strong additivity 
in \cite[Theorem E]{W} also has a similar trouble, but the arguments in
\cite{TL} gives a correct proof of the strong additivity of the
$SU(n)_k$-net and the results in \cite{W} are not affected.)
A. Wassermann informed us that he can fix this error and recover the
results in \cite{Lk}.  (Note that the strong additivity for $\Vir_c$ 
with $c<1$ follows from our Corollary \ref{coset-rat}.)
If we can use the results in \cite{Lk} directly,
we can give an alternate proof of the results in this section as
follows.  First, Loke's results imply that the Virasoro nets are
rational in the sense that we have only finitely many irreducible
DHR endomorphisms and that all of them have finite indices.  This is
enough for showing that the Virasoro net with $c<1$ is contained in the
corresponding coset net irreducibly as in the remark after the proof of
Proposition \ref{Vir-sub}.  Then Proposition \ref{finite-index} implies
that the index is finite and this already shows that the Virasoro net
is completely rational by \cite{L4}.  Then by comparing the $\mu$-indices
of the Virasoro net and the coset net, we conclude that
the two nets are equal.
}\end{remark}

\section{Classification of local extensions of the Virasoro nets}
\label{class-ext-Virasoro}

By \cite{CIZ}, we have a complete classification of the modular invariants
for the Virasoro nets with
central charge $c=1-6/m(m+1)<1$, $m=2,3,4,\dots$.
If each modular invariant is realized with $\a$-induction
for an extension $\Vir_c\subset \M$ as in \cite[Corollary 5.8]{BEK1},
then we  have the numbers of
irreducible morphisms as in Tables \ref{min-I}, \ref{min-II} by
a similar method to the one used in \cite[Table 1, page 774]{BEK2},
where $|{}_\N \Delta_\M|$, $|{}_\M \Delta_\M|$,
$|{}_\M \Delta^+_\M|$, and $|{}_\M \Delta^0_\M|$ denote the numbers
of irreducible $\N$-$\M$ sectors, $\M$-$\M$ sectors, $\M$-$\M$ sectors
arising from $\a^\pm$-induction, and the ambichiral $\M$-$\M$ sectors,
respectively.  (The ambichiral sectors are those arising from both
$\a^+$- and $\a^-$-induction, as in \cite[page 741]{BEK2}.)
We will prove that the entries in Table \ref{min-I} correspond bijectively
to local extensions of the Virasoro nets and that each entry
in Table \ref{min-II} is realized with a non-local extension of
the Virasoro net.  (For the labels for $Z$ in Table \ref{min-I},
see Table \ref{Vir-mod-I}.)

\begin{table}[htbp]
\begin{center}{\footnotesize
\begin{tabular}{|c|c|c|c|c|c|}\hline
$m$ & Labels for $Z$ & $|{}_\N \Delta_\M|$ & $|{}_\M \Delta_\M|$
& $|{}_\M \Delta^+_\M|$ & $|{}_\M \Delta^0_\M|$ \\ \hline
$n$ & $(A_{n-1},A_n)$ & $n(n-1)/2$
& $n(n-1)/2$ & $n(n-1)/2$ & $n(n-1)/2$ \\ \hline
$4n+1$ & $(A_{4n},D_{2n+2})$ & $2n(2n+2)$
& $2n(4n+4)$ & $2n(2n+2)$ & $2n(n+2)$ \\ \hline
$4n+2$ & $(D_{2n+2},A_{4n+2})$ & $(2n+1)(2n+2)$
& $(2n+1)(4n+4)$ & $(2n+1)(2n+2)$ & $(2n+1)(n+2)$ \\ \hline
11  & $(A_{10},E_6)$ & 30 & 60 & 30 & 15  \\ \hline
12  & $(E_6, A_{12})$ & 36 & 72 & 36 & 18  \\ \hline
29  & $(A_{28},E_8)$ & 112 & 448 & 112 & 28  \\ \hline
30  & $(E_8, A_{30})$ & 120 & 480 & 120 & 30  \\ \hline
\end{tabular}}
\caption{Type I modular invariants for the Virasoro nets}
\label{min-I}
\end{center}
\end{table}

\begin{table}[htbp]
\begin{center}{\footnotesize
\begin{tabular}{|c|c|c|c|c|c|}\hline
$m$ & Labels for $Z$ & $|{}_\N \Delta_\M|$ & $|{}_\M \Delta_\M|$
& $|{}_\M \Delta^+_\M|$ & $|{}_\M \Delta^0_\M|$ \\ \hline
$4n$ & $(D_{2n+1}, A_{4n})$ & $2n(2n+1)$
& $2n(4n-1)$ & $2n(4n-1)$ & $2n(4n-1)$ \\ \hline
$4n+3$ & $(A_{4n+2},D_{2n+3})$ & $(2n+1)(2n+3)$
& $(2n+1)(4n+3)$ & $(2n+1)(4n+3)$ & $(2n+1)(4n+3)$ \\ \hline
17  & $(A_{16},E_7)$ & 56 & 136 & 80 & 48  \\ \hline
18  & $(E_7, A_{18})$ & 63 & 153 & 90 & 54  \\ \hline
\end{tabular}}
\caption{Type II modular invariants for the Virasoro nets}
\label{min-II}
\end{center}
\end{table}

\begin{theorem}
\label{Vir-ext}
The local irreducible extensions of the Virasoro nets on the circle
with central charge less than 1 correspond bijectively to
the entries in Table \ref{min-I}.
\end{theorem}

Note that the index $[\M:\N]$ in the seven cases in Table \ref{min-I}
are 1, 2, 2, $3+\sqrt3$, $3+\sqrt3$,
$\sqrt{30-6\sqrt{5}}/2\sin(\pi/30)=19.479\cdots$,
$\sqrt{30-6\sqrt{5}}/2\sin(\pi/30)=19.479\cdots$, respectively.

\begin{theorem}
\label{Vir-ext2}
Each entry in Table \ref{min-II} is realized by $\a$-induction for
a non-local (but relatively local) extension of the Virasoro net 
with central charge $c=1-6/m(m+1)$.
\end{theorem}

Proofs of these theorems are given in the following subsections. 
 
\begin{remark}{\rm We make here explicit that every irreducible net 
extension $\A$ of $\Vir_c$, $c<1$, is diffeomorphism covariant.  

First note that every representation $\rho$ of 
$\Vir_c$ is diffeomorphism covariant; indeed we can assume that 
$d(\rho)<\infty$ (by decomposition into irreducibles) thus $\rho$ is 
M\"{o}bius covariant with positive energy by \cite{GL1} because
$\Vir_c$ is strongly additive. Then $\rho$ is diffeomorphism covariant 
by Lemma \ref{diffeo1}. 

Now fix an interval $I\subset S^1$ 
and consider a canonical endomorphism $\gamma_I$ of $\A(I)$ into 
$\Vir_c(I)$ so that $\th_I\equiv \gamma_I 
\upharpoonright_{\Vir_c(I)}$ is the restriction of a DHR endomorphism $\th$
localized in $I$. With $z_{\th}$ the covariance cocycle of 
$\th$, the covariant action of $\Diff(S^1)$ on $\A$ is given by 
\[
\tilde\a_g(X)=\a_g(X),\qquad \tilde\a_g(T)=z_{\th}(g)^*T, \qquad g\in\Diff(S^1)
\]
where $X$ is a local operator of $\Vir_c$, $T\in\A(I)$ is isometry 
intertwining the identity and $\gamma_I$ and $\a$ is the covariant 
action of $\Diff(S^1)$ on $\Vir_c$ (cf. \cite{L4}).
}\end{remark}

\subsection{Simple current extensions}

First we handle the easier case, the simple current extensions of index 2
in Theorem \ref{Vir-ext2}.

Let $\N$ be the Virasoro net
with central charge $c=1-6/m(m+1)$.  We have irreducible DHR
endomorphisms $\la_{(p,q)}$ as in Subsection \ref{Vir-alg-net}.
The statistics phase of the sector $\la_{(m-1,1)}$ is
$\exp (\pi i(m-1)(m-2)/2)$ by (\ref{spin}).  This is
equal to 1 if $m\equiv 1,2 \mod 4$, and $-1$
if $m\equiv 0,3\mod 4$.  In both cases, we can take an automorphism
$\sigma$ with $\sigma^2=1$ within the unitary equivalence class 
of the sector $[\la_{(m-1,1)}]$ by \cite[Lemma 4.4]{R2}.
It is clear that $\rho=\id\oplus\sigma$ is an endomorphism
of a $Q$-system, so we can make an irreducible extension $\M$ with 
index 2 by \cite[Theorem 4.9]{LR}.  By \cite[II, Corollary 3.7]{BE}, 
the extension is local if and only if $m\equiv 1,2 \mod 4$.
The extensions are unique for each $m$, because of
triviality of $H^2(\Z/2\Z, \T)$ and \cite{IK}, and we get the modular 
invariants as in Tables \ref{min-I}, \ref{min-II}.
(See \cite[II, Section 3]{BE} for similar computations.)

\subsection{The four exceptional cases}

We next handle the remaining four exceptional cases in
Theorem \ref{Vir-ext2}.

We first deal with the case $m=11$ for the modular invariants
$(A_{10},E_6)$.
The other three cases can be handled in very similar ways.

Let $\N$ be the Virasoro net with central charge $c=21/22$.
Fix an interval $I$ on the circle and 
consider the set of DHR endomorphisms of the net $\N$
localized in $I$ as in Subsection \ref{Vir-alg-net}.
Then consider the subset $\{\la_{(1,1)},\la_{(1,2)},\dots,\la_{(1,11)}\}$
of the DHR endomorphisms.
By the fusion rules (\ref{fusion}), this system is closed under composition
and conjugation, and the fusion rules are the same as for
$SU(2)_{10}$.  So the subfactor $\la_{(1,2)}(\N(I))\subset \N(I)$
has the principal graph $A_{11}$ and the fusion rules and the
quantum $6j$-symbols for the subsystem
$\{\la_{(1,1)},\la_{(1,3)},\la_{(1,5)},\dots,\la_{(1,11)}\}$
of the DHR endomorphisms are the same as those for the usual
Jones subfactor with principal graph $A_{11}$ and uniquely
determined. (See \cite{O1}, \cite{K1},
\cite[Chapters 9--12]{EK}.) 
Since we already know by Theorem \ref{class-SU2} that
the endomorphism $\la_0\oplus\la_6$ gives a $Q$-system uniquely
for the system of irreducible DHR sectors
$\{\la_0,\la_1,\dots,\la_{10}\}$ for the $SU(2)_{10}$ net,
we also know that the endomorphism $\la_{(1,1)}\oplus\la_{(1,7)}$
gives a $Q$-system uniquely, by the above identification of
the fusion rules and quantum $6j$-symbols.  By \cite[Theorem 4.9]{LR},
we can make an irreducible extension $\M$ of $\N$ using this
$Q$-system, but the locality criterion in \cite[Theorem 4.9]{LR}
depends on the braiding structure of the system, and the
standard braiding on the $SU(2)_{10}$ net and the braiding
we know have on 
$\{\la_{(1,1)},\la_{(1,2)},\dots,\la_{(1,11)}\}$ from the Virasoro
net are not the same, since their spins are different.  So we need
an extra argument for showing the locality of the extension.

Even when the extension
is not local, we can apply the $\a$-induction to the subfactor
$\N(I)\subset \M(I)$ and then the matrix $Z$ given by
$Z_{\la\,\mu}=\lan \a^+_\la,\a^-_\mu\ran$
is a modular invariant for the $S$ and $T$ matrices arising
from the minimal model by \cite[Corollary 5.8]{BEK1}.  (Recall that
the braiding is now non-degenerate.)
By the Cappelli-Itzykson-Zuber classification \cite{CIZ}, we have only
three possibilities for this matrix at $m=11$.  It is now easy to
count the number of $\N(I)$-$\M(I)$ sectors arising from all
the DHR sectors of $\N$ and the embedding $\iota:
\N(I)\subset \M(I)$ as in \cite{BEK1,BEK2}, and
the number is 30.  Then by \cite{BEK1} and the Tables 
\ref{min-I}, \ref{min-II},
we conclude that the matrix $Z$ is of type $(A_{10}, E_6)$.
Then by a criterion of locality due to B\"ockenhauer-Evans
\cite[Proposition 3.2]{BE4}, we conclude from this modular invariant
matrix that the extension $\M$ is local.  The uniqueness of
$\M$ also follows from the above argument.
(Uniqueness in Theorem \ref{class-SU2} is under an assumption of
locality, but the above argument based on \cite{BE4} shows that
an extension is automatically local in this setting.)

In the case of $m=12$ for the modular invariant $(E_6, A_{12})$,
we now use the system
$\{\la_{(1,1)},\la_{(2,1)},\dots,\la_{(11,1)}\}$.  Then the
rest of the arguments are the same as above.
The cases $m=29$ for the modular invariant $(A_{28},E_8)$
and $m=30$ for the modular invariant $(E_8, A_{30})$ are handled
in similar ways.

\begin{remark}{\rm
In the above cases, we can determine the isomorphism class of the 
subfactors $\N(I)\subset \M(I)$ for a fixed interval $I$ as follows.
Let $m=11$.  By the same arguments as in \cite[Appendix]{BEK2},
we conclude that the subfactor $\N(I)\subset \M(I)$ is the Goodman-de 
la Harpe-Jones subfactor \cite[Section 4.5]{GHJ} of index $3+\sqrt3$
arising from the Dynkin diagram $E_6$.  We get the isomorphic
subfactor also for $m=12$.  The cases $m=29,30$ give the
Goodman-de la Harpe-Jones subfactor arising from $E_8$.
}\end{remark}

\subsection{Non-local extensions}

We now explain how to prove Theorem \ref{Vir-ext2}.
We have already seen the case of $D_{\text{odd}}$ above. 
In the case of $m=17,18$ for the modular invariants
of type $(A_{16}, E_7)$, $(E_7, A_{18})$, respectively,
we can make $Q$-systems in very similar ways to the above
cases.  Then we can make the extensions $\M(I)$, but
the criterion in \cite[Proposition 3.2]{BE4} shows that
they are not local. The extensions are relatively 
local by \cite[Th. 4.9]{LR}.

\subsection{The case $c=1$}

By \cite{R3}, we know that the Virasoro net for $c=1$ is
the fixed point net of the $SU(2)_1$ net with the action of
$SU(2)$.
That is, for each closed subgroup of $SU(2)$, we  have a fixed
point net, which is an irreducible local
extension of the Virasoro net with $c=1$.
Such subgroups are labeled with affine $A$-$D$-$E$ diagrams
and we have infinitely many such subgroups.
(See \cite[Section 4.7.d]{GHJ}, for example.)  Thus finiteness
of local extensions fails for the case $c=1$.

Note also that, if $c>1$, $\Vir_c$ is not strongly additive \cite{BS} 
and all sectors, but the identity, are expected to be 
infinite-dimensional  \cite{R3}.

\section{Classification of conformal nets}
\label{class-diffeo}

We now give our main result.

\begin{theorem}
\label{diffeo}
The local (irreducible) conformal nets on the circle
with central charge less than 1 correspond bijectively to
the entries in Table \ref{min-I}.
\end{theorem}

\begin{proof}
By Proposition \ref{Vir-sub}, a conformal net
$\M$ on the circle with central charge less than 1
contains a Virasoro net as an irreducible subnet.
Thus Theorem \ref{Vir-ext} gives the desired conclusion.
\end{proof}

In this theorem, the correspondence between such conformal
nets and pairs of Dynkin diagrams is given explicitly as follows.
Let $\M$ be such a net with central charge $c<1$ and $\Vir_c$ 
its canonical Virasoro subnet as above.  Fix an interval $I\subset S^1$.
For a DHR endomorphism $\la{(p,q)}$ of $\Vir_c$ localized in $I$, we have
$\a^\pm$-induced endomorphism $\a^\pm_{\la{(p,q)}}$ of $\M(I)$.  We
denote this endomorphism simply by $\a^\pm_{(p,q)}$.  Then
we have two subfactors $\a^+_{(2,1)}(\M(I))\subset \M(I)$ and
$\a^+_{(1,2)}(\M(I))\subset \M(I)$ and the index values are both below
4.  Let $(G, G')$ be the pair of the corresponding principal graphs
of these two subfactors.  The above main theorem says that the map
from $\M$ to $(G,G')$ gives a bijection from the set of isomorphism
classes of such nets to the set of pairs $(G,G')$ of
$A_n$-$D_{2n}$-$E_{6,8}$ Dynkin diagrams such that the Coxeter 
number of $G$ is smaller than that of $G'$ by 1.

\section{Applications and remarks}
\label{appl}

In this section, we identify some coset nets studied in \cite{BE,X6}
in our classification list, as applications of our main results.

\subsection{Certain coset nets and extensions of the Virasoro nets}

In \cite[Section 3.7]{X6}, Xu considered the three coset nets 
arising from
$SU(2)_8\subset SU(3)_2$, $SU(3)_2\subset SU(3)_1\times SU(3)_1$,
$U(1)_6\subset SU(2)_3$, all at central charge $4/5$.  He found that
all have six simple objects in the tensor categories of the DHR
endomorphisms and give the same invariants for 3-manifolds. 
Our classification theorem \ref{diffeo} shows that these three
nets are indeed isomorphic as follows.

Theorem \ref{diffeo} shows that we have only two conformal
nets at central charge $4/5$.  One is the Virasoro net itself
with $m=5$ that has 10 irreducible DHR endomorphisms, and the other
is its simple current extension of index 2 that has 6 irreducible
DHR endomorphisms.  This implies that all the three cosets above
are isomorphic to the latter.
\subsection{More coset nets and extensions of the Virasoro nets}

For the local extensions of the Virasoro nets corresponding
to the modular invariants $(E_6, A_{12})$, $(E_8, A_{30})$,
B\"ockenhauer-Evans \cite[II, Subsection 5.2]{BE} say that
``the natural candidates'' are the cosets arising from
$SU(2)_{11}\subset SO(5)_1\times SU(2)_1$ and
$SU(2)_{29}\subset (G_2)_1\times SU(2)_1$, respectively, but
they were unable to prove that these cosets indeed produce
the desired local extensions.  (For the modular invariants
$(A_{10}, E_6)$, $(A_{28}, E_8)$, they also say that ``there
is no such natural candidate'' in \cite[II, Subsection 5.2]{BE}.)
It is obvious that the above
two cosets give local irreducible extensions of the Virasoro
nets, but the problem is that the index might be 1.  Here
we already have a complete classification of local irreducible
extensions of the Virasoro nets, and using it, we can prove
that the above two cosets indeed coincide with the extension
we have constructed above.

First we consider the case of the modular invariant $(E_6, A_{12})$.
Let $\A$, $\B$, ${\cal C}$ be the nets corresponding
to $SU(2)_{11}$, $SU(2)_{10}\times SU(2)_1$, $SO(5)_1 \times SU(2)_1$,
respectively.  We have natural inclusions $\A(I)\subset \B(I) \subset
{\cal C}(I)$, and define the coset nets by $\D(I)=\A(I)'\cap \B(I)$,
$\E(I)=\A(I)'\cap {\cal C}(I)$.  We know that the net $\D(I)$ is the
Virasoro net with central charge $25/26$ and will prove that
the extension $\E$ is the one corresponding to the entry
$(E_6, A_{12})$ in Table \ref{min-I} in Theorem \ref{Vir-ext}.

The following diagram 
$$\begin{array}{ccc}
\label{comm-sq}
\A(I) \vee \D(I) & \subset & \B(I) \\
\cap && \cap \\
\A(I) \vee \E(I) &\subset &{\cal C}(I)
\end{array}$$
is a commuting square \cite{P1}, \cite[Chapter 4]{GHJ}, and we have
\begin{equation}\label{ineq}
[\B(I): \A(I) \vee \D(I)] \le [{\cal C}(I): \A(I) \vee \E(I)] < \infty.
\end{equation}
Next note that the new coset net $\{\E(I)'\cap {\cal C}(I)\}$ gives
an irreducible local extension of the net $\A$, but
Theorem \ref{class-SU2} implies that we have no strict
extension of $\A$.  Thus we have $\E(I)'\cap {\cal C}(I)=\A(I)$,
and $\A(I), \E(I)$ are the relative commutants of each other
in ${\cal C}(I)$.  So we can consider the inclusion
$\A(I)\otimes \E(I)\subset {\cal C}(I)$ and this is a canonical
tensor product subfactor in the sense of Rehren \cite{R4,R5}.
(See \cite[line 22--24 in page 701]{R4}.)
Thus the dual canonical endomorphism for this subfactor
is of the form $\bigoplus_j \sigma_j\otimes\pi(\sigma_j)$,
where $\{\sigma_j\}$ is a closed subsystem of DHR endomorphisms
of the net $\A$ and the map $\pi$ is a bijection from
this subsystem to a closed subsystem of DHR endomorphisms
of the net $\E$, by \cite[Corollary 3.5, line 3--12 in page 706]{R4}.
This implies that the index   $[{\cal C}(I): \A(I) \vee \E(I)]$ is
a square sum of the statistical dimensions of the irreducible
DHR endomorphisms over a subsystem of the $SU(2)_{11}$-system.
We have only three possibilities for such a closed subsystem
as follows.
\begin{enumerate}
\item $\{\la_0=\id\}$
\item The even part $\{\la_0,\la_2,\dots,\la_{10}\}$
\item The entire system  $\{\la_0,\la_1,\dots,\la_{11}\}$
\end{enumerate}
The first case would violate the inequality (\ref{ineq}).
Recall that we have only two possibilities for $\mu_{\E}$
by Theorem \ref{Vir-ext} and that
we also have equality
\begin{equation}\label{eq}
\mu_\A \mu_\E =\mu_{\cal C} [{\cal C}(I): \A(I) \vee \E(I)]
\end{equation}
by \cite[Proposition 24]{KLM}.
Then the third case of the above three would be incompatible
with the above equality (\ref{eq}), and
thus we conclude that the second case occurs.  Then the above
equality (\ref{eq}) easily shows that
the extension $\E(I)$ is the one corresponding to the entry
$(E_6, A_{12})$ in Table \ref{min-I} in Theorem \ref{Vir-ext}.

The case $(E_8, A_{30})$ can be proved with a very similar
argument to the above.  We now have three possibilities
for the $\mu$-index by Theorem \ref{Vir-ext} instead of
two possibilities above, but this causes no problem, and
we get the desired isomorphism.

\subsection{Subnet structure}

As a consequence of our results, the subnet structure of a local 
conformal net with $c<1$ is very simple. 

Let  $\A$ be a local 
irreducible conformal net on $S^1$ with $c<1$. The projective unitary 
representation $U$ of $\Diff(S^1)$ is given so the central charge
and the Virasoro subnet are well-defined. By our classification, the 
Virasoro subnet (up to conjugacy), thus the central charge, do not 
depend on the choice of the covariance representation $U$ if $c<1$.

The following elementary lemma is implicit in the literature.

\begin{lemma}\label{ufd}
Every projective unitary finite-dimensional representation of $\Diff(S^1)$ is 
trivial.
\end{lemma}

\begin{proof}
Otherwise, passing to the infinitesimal representation, we have 
operators $L_n$ and $c$ on a finite-dimensional Hilbert space 
satisfying the Virasoro relations (\ref{vir-rel})
and the unitarity conditions $L_n^*=L_{-n}$. Then $\{L_1,L_{-1},L_0\}$ 
gives a unitary finite-dimensional representation of the Lie algebra 
$s\ell(2,\mathbb R)$, thus $L_1 = L_{-1} = L_0 = 0$. Then for $m\neq 
0$ we have $L_m=m^{-1}[L_m,L_0]= 0$ and also $c=0$ due to the 
relations (\ref{vir-rel}).
\end{proof}

\begin{proposition}\label{ba}
Let $\A$ be a local conformal net and $\B\subset\A$ a conformal subnet 
with finite index.  
Then $\B$ contains the Virasoro subnet:
$\B(I)\supset \A_{\Vir}(I)$, $I\in\I$.
\end{proposition}

\begin{proof}
Let $\pi_0$ denote the vacuum representation of $\A$. 
As $[\A:\B]<\infty$ we have an irreducible decomposition
\begin{equation}\label{ve}
\pi_0 |_{\B} = \bigoplus_{i=0}^n n_i \r_i ,
\end{equation}
with $n_i<\infty$. Accordingly the vacuum Hilbert space $\H$ of $\A$ 
decomposes as $\H=\bigoplus_i \H_i\otimes\K_i$ where dim$\K_i = n_i$.

By assumptions the projective unitary representation $U$ implements 
automorphisms of $\pi_0(\B)''$, hence of its commutant 
$\pi_0(\B)'\simeq \bigoplus_i 1|_{\H_i}\otimes B(\K_i)$ 
which is finite-dimensional. As $\Diff(S^1)$ is 
connected, Ad$U$ acts trivially on the center of $\pi_0(\B)'$, hence 
it implements automorphisms on each simple summand of $\pi_0(\B)'$, 
isomorphic to $B(\K_i)$, hence it gives rise to a 
finite-dimensional representation of $\Diff(S^1)$ that is unitary with 
respect to the tracial scalar product, and so must be trivial because 
of Lemma \ref{ufd}.
It follows that $U$ decomposes according to eq. (\ref{ve}) as
\[
U= \bigoplus_{i=0}^n  U_i\otimes 1|_{\K_i} 
\]
where $U_i$ is a covariance representation for $\r_i$. Thus 
$U(\Diff(I))\subset \bigoplus_i B(\H_i)\otimes 1|_{\K_i}=\pi_0(\B)''$, 
so $\A_{\Vir}(I)\subset \pi_0(\B)'\cap\A(I)$ which equals $\B(I)$ 
by Lemma \ref{subnet}.
\end{proof}

\begin{theorem} Let $\A$ be an irreducible local conformal net with 
central charge $c<1$. Let $s$ be the number of 
finite-index conformal subnets, up to conjugacy (including $\A$ itself). Then 
$s\in\{1,2,3\}$. $\A$ is completely classified by the 
pair $(m,s)$ where $c=1-6/m(m+1)$. For any $m\in\mathbb N$ the 
possible values of $s$ are: 
\begin{enumerate}
\item $s=1$ for all $m\in\mathbb N$;
\item $s=2$ if $m=1,2$ {\rm mod} $4$, and if $m=11,12$;
\item $s=3$ if $m=29,30$.
\end{enumerate}
The corresponding structure follows from Table \ref{min-I}.
\end{theorem}

\begin{proof} The proof is immediate by the classification Theorem 
\ref{diffeo} and Proposition \ref{ba}.
\end{proof}

\subsection{Remarks on subfactors and commuting squares}
\label{rema}

It is interesting to point out that our 
framework of nets of subfactors as in \cite{LR}
can be regarded as a net version of the usual classification
problem of subfactors \cite{J}. The difference here is that the
smaller net is fixed and we wish to classify extensions, while in
the usual subfactor setting a larger factor is fixed and we would
like to classify factors contained in it.  In the subfactor theory,
classifying subfactors and classifying extensions are equivalent 
problems because of Jones basic construction \cite{J} 
(as long as we have finite index),
but this is not true in the setting of nets of subfactors.  Here, 
the basic construction does not work and considering
an extension and considering a subnet are not symmetric procedures.
(For a net of subfactors
$\N\subset \M$, the dual canonical endomorphism for
$\N(I)\subset \M(I)$ decomposes into DHR endomorphisms of the net
$\N$, but the canonical endomorphism for $\N(I)\subset \M(I)$
does not decompose into  DHR endomorphisms of the net $\M$.)

To illustrate this point, consider the example of a completely
rational net $SU(2)_1$.  This net has an action of
$SU(2)$ by internal symmetries, so a fixed point subnet with respect to
any finite subgroup of $SU(2)$. We have infinitely many
such finite subgroups,  thus the completely rational
net $SU(2)_1$ has infinitely many irreducible subnets with finite index.
On the other hand, the number of irreducible
extensions of a given completely rational
net is always finite, since the number of mutually 
inequivalent $Q$-systems $(\rho, V, W)$ is finite for a given 
$\rho$ by \cite{IK} and
we have only finitely many choices of $\rho$ for a given
completely rational net,
and this finite number is often very small, as shown in the main body
of this paper.  In general, considering
extensions gives much stronger constraints than considering subnets,
and this allows an interesting classification in concrete models.

Notice now that a net of factors on the circle
produces a tensor category of DHR endomorphisms.  
On the other hand a subfactor $N\subset M$ with
finite index produces tensor categories of endomorphisms of $N$ and $M$
arising from the powers of (dual) canonical endomorphisms.  
In this analogy, complete rationality corresponds to 
the finite depth condition for subfactors,
and the 2-interval inclusion has similarity to the 
construction in \cite{LR},
or the quantum double construction, as explained in \cite{KLM}.
A net of subfactors corresponds to ``an inclusion
of one subfactor into another subfactor'', that is,
a commuting square of factors \cite{P1}, studied in \cite{K2}.
For any subfactor $N\subset M$ with finite index, we have a Jones
subfactor $P\subset Q$ made of the Jones projections
with same index \cite{J} such that we have a commuting square
$$\begin{array}{ccc}
N & \subset & M \\
\cup && \cup \\
P & \subset & Q.
\end{array}$$
In this sense, the Jones subfactors are ``minimal'' among general
subfactors.  The Virasoro nets have a similar minimality among nets
of factors with diffeomorphism covariance, 
they are contained in every local 
conformal net (but they do not admit any non-trivial subnet \cite{C}).
This similarity is a guide to understanding our work.

In the above example of a commuting
square, we have no control over an inclusion $P\subset N$ in general, but
in the case of Virasoro net, we do have a control over the inclusion
if the central charge is less than 1.  This has enabled us to obtain our
results.  As often pointed out, the
condition that the Jones index is less than 4 has some formal
similarity to the condition that the central charge is less than 1.
The results in this paper give further evidence for this similarity.

\medskip
\noindent{\bf Acknowledgments.}
A part of this work was done during a visit of the first-named
author to Universit\`a di Roma ``Tor Vergata''.
We gratefully acknowledge the financial support of 
GNAMPA-INDAM and MIUR (Italy) and 
Grants-in-Aid for Scientific Research, JSPS (Japan).
We thank S. Carpi, M. M\"uger, K.-H. Rehren
and F. Xu for answering our questions.  We are grateful to
M. Izumi for explaining the result in \cite{IK} and 
a criticism to our original manuscript.
We also thank A. Wassermann for informing us of a mistake
in \cite{Lk}.


{\footnotesize \begin{thebibliography}{99}

\bibitem{AH}
M. Asaeda \& U. Haagerup,
{\it Exotic subfactors of finite depth with Jones indices
${(5+\sqrt{13})}/{2}$ and ${(5+\sqrt{17})}/{2}$},
Commun. Math. Phys. {\bf 202} (1999) 1--63.

\bibitem{BPZ}
A. A. Belavin, A. M. Polyakov \& A. B. Zamolodchikov, 
{\it Infinite conformal symmetry in two-dimensional quantum field theory}, 
Nucl. Phys. {\bf 241} (1984) 333--380.

\bibitem{BE}
J. B\"ockenhauer \& D. E. Evans,
{\it Modular invariants, graphs and $\alpha$-induction for
nets of subfactors I},
Commun. Math. Phys. {\bf 197} (1998) 361--386. II
{\bf 200} (1999) 57--103. III {\bf 205} (1999) 183--228.

\bibitem{BE4}
J. B\"ockenhauer \& D. E. Evans, 
{\it Modular invariants from subfactors: Type I coupling matrices and
intermediate subfactors}, 
Commun. Math. Phys. {\bf 213} (2000) 267--289.

\bibitem{BEK1}
J. B\"ockenhauer, D. E. Evans \& Y. Kawahigashi,
{\it On $\a$-induction, chiral projectors
and modular invariants for subfactors},
Commun. Math. Phys.  {\bf 208} (1999) 429--487.

\bibitem{BEK2}
J. B\"ockenhauer, D. E. Evans \& Y. Kawahigashi,
{\it Chiral structure of modular
invariants for subfactors}, 
Commun. Math. Phys. {\bf 210} (2000) 733--784.

\bibitem{BEK3}
J. B\"ockenhauer, D. E. Evans \& Y. Kawahigashi,
{\it Longo-Rehren subfactors arising from $\alpha$-induction},
Publ. RIMS, Kyoto Univ. {\bf 37} (2001) 1--35.

\bibitem{BGL} R. Brunetti, D. Guido \& R. Longo,
{\it Modular structure and duality in conformal
quantum field theory}, Commun. Math. Phys.
{\bf 156} (1993) 201--219.

\bibitem{BMT} D. Buchholz, G. Mack \& I. Todorov,
{\it The current algebra on the circle as a germ of local field theories}, 
Nucl. Phys. B, Proc. Suppl. {\bf 5B} (1988) 20--56.

\bibitem{BS} D. Buchholz \& H. Schulz-Mirbach,
{\it Haag duality in conformal quantum field theory}, Rev. Math. 
Phys. {\bf 2} (1990) 105--125.

\bibitem{CIZ}
A. Cappelli, C. Itzykson \& J.-B. Zuber,
{\it The $A$-$D$-$E$ classification of minimal and
$A^{(1)}_1$ conformal invariant theories},
Commun. Math. Phys.  {\bf 113}  (1987) 1--26.

\bibitem{C} S. Carpi, {\it Absence of subsystems for the Haag-Kastler 
net generated by the energy-momentum tensor in two-dimensional conformal 
field theory}, Lett. Math. Phys. {\bf 45} (1998) 259--267.

\bibitem{DRL} C. D'Antoni,  R. Longo \& F. Radulescu,
{\it Conformal nets, maximal temperature and models from free
probability}, J. Operator Theory {\bf 45} (2001) 195--208.

\bibitem{DMS}
P. Di Francesco, P. Mathieu \& D. S\'en\'echal,
``Conformal Field Theory'', Springer-Verlag, 
Berlin-Heidelberg-New York, 1996.

\bibitem{DHR}   S. Doplicher, R. Haag \& J. E. Roberts,
{\it Local observables and particle statistics}, I. Commun. Math. Phys.
{\bf 23} (1971) 199--230; II. {\bf 35} (1974) 49--85.

\bibitem{DL}  S. Doplicher \& R. Longo,
{\it Standard and split inclusions of von Neumann algebras},
Invent. Math.  {\bf 73} (1984) 493--536.

\bibitem{EK} D. E. Evans \& Y. Kawahigashi,
``Quantum Symmetries on Operator Algebras'',
Oxford University Press, Oxford, 1998.

\bibitem{FJ} K.  Fredenhagen \& M. J\"or\ss, 
{\it Conformal Haag-Kastler nets, pointlike localized fields and the 
existence of operator product expansion}, 
Commun. Math. Phys.  {\bf 176} (1996) 541--554.

\bibitem{FRS}
K. Fredenhagen, K.-H. Rehren \& B. Schroer,
{\it Superselection sectors with braid group statistics
and exchange algebras}, 
I. Commun. Math. Phys.  {\bf 125} (1989) 201--226,
II. Rev. Math. Phys. {\bf Special issue} (1992) 113--157.

\bibitem{FQS}
D. Friedan, Z. Qiu \& S. Shenker,  
{\it Details of the non-unitarity proof for highest weight 
representations of the Virasoro algebra}, 
Commun. Math. Phys. {\bf 107} (1986) 535--542.

\bibitem{FG} 
J. Fr\"ohlich \& F. Gabbiani,
{\it Operator algebras and conformal field theory},
Commun.  Math.  Phys.  {\bf 155} (1993) 569--640.

\bibitem{G}
T. Gannon,
{\it Modular data: the algebraic combinatorics of
conformal field theory}, preprint 2001, math.QA/0103044.

\bibitem{GKO}
P. Goddard, A. Kent \& D. Olive,
{\it Unitary representations of the Virasoro and
super-Virasoro algebras},
Commun.  Math.  Phys. {\bf 103} (1986) 105--119.

\bibitem{GHJ}
F. Goodman, P. de la Harpe \& V. F. R. Jones,
``Coxeter Graphs and Towers of Algebras''
MSRI Publications, Springer, Berlin-Heidelberg-New York,
{\bf 14}, 1989.

\bibitem{GL1}
D. Guido \& R. Longo, 
{\it Relativistic invariance and
charge conjugation in quantum field theory}, 
Commun.  Math.  Phys.  {\bf 148} (1992) 521---551.

\bibitem{GL2}
D. Guido \& R. Longo, {\it The conformal spin and statistics theorem},
Commun. Math.  Phys.  {\bf 181} (1996) 11--35.

\bibitem{GLW}
D. Guido, R. Longo \& H.-W. Wiesbrock, {\it Extensions of conformal
nets and superselection structures}, Commun.  Math.  Phys.  {\bf 192}
(1998) 217--244.

\bibitem{HL}
P. Hislop \& R. Longo, {\it Modular structure of the local algebras
associated with the free massless scalar field theory},
Commun. Math. Phys. {\bf 84} (1982) 71--85.

\bibitem{H} R. Haag ``Local Quantum Physics", Springer-Verlag,
Berlin-Heidelberg-New York, (1996).

\bibitem{I1} M. Izumi,
{\it Subalgebras of infinite $C^*$-algebras with
finite Watatani indices II: Cuntz-Krieger algebras},
Duke Math. J. {\bf 91} (1998) 409--461.

\bibitem{I2} M. Izumi,
{\it The structure of sectors associated with the Longo-Rehren
inclusions}, Commun. Math. Phys. {\bf 213} (2000) 127--179.

\bibitem{IK} M. Izumi \& H. Kosaki, 
{\it On a subfactor analogue of the second cohomology},
Rev. Math. Phys. {\bf 14} (2002) 733--757.

\bibitem{ILP} M. Izumi, R. Longo \& S. Popa
{\it  A Galois correspondence for compact groups of automorphisms
of von Neumann algebras with a generalization to Kac algebras},
J. Funct. Anal. {\bf 10} (1998) 25--63.

\bibitem{J}
V. F. R. Jones, {\it Index for subfactors}, Invent.  Math.  {\bf 72}
(1983) 1--25.

\bibitem{KR}
V. G. Kac \& A. K. Raina, 
``Highest Weight Representations of Infinite Dimensional Lie 
Algebras'', World Scientific 1987.

\bibitem{Kt}
A. Kato, 
{\it Classification of modular invariant partition
functions in two dimensions},
Modern Phys. Lett {\bf A2} (1987) 585--600.

\bibitem{K1}
Y. Kawahigashi, 
{\it On flatness of Ocneanu's connections on the Dynkin diagrams
and classification of subfactors},
J. Funct. Anal. {\bf 127} (1995) 63--107.

\bibitem{K2}
Y. Kawahigashi,
{\it Classification of paragroup actions on subfactors},
Publ. RIMS, Kyoto Univ. {\bf 31} (1995) 481--517.

\bibitem{KL}
Y. Kawahigashi \& R. Longo,
{\it Classification of two-dimensional local conformal nets with $c<1$
and 2-cohomology vanishing for tensor categories}, preprint.

\bibitem{KLM}
Y. Kawahigashi, R. Longo \& M. M\"uger, 
{\it Multi-interval subfactors and modularity of representations
in conformal field theory}, Commun. Math. Phys.
{\bf 219} (2001) 631--669.

\bibitem{KO}
A. Kirillov Jr. \& V. Ostrik,
{\it On $q$-analog of McKay correspondence and ADE classification of
$sl^{(2)}$ conformal field theories},
Adv. Math. {\bf 171} (2002) 183--227.

\bibitem{Lk}
T. Loke,
{\it Operator algebras and conformal field theory of the
discrete series representations of {\rm Diff}$(S^1)$},
Thesis, University of Cambridge, 1994.

\bibitem{L1} R. Longo,
{\it Index of subfactors and statistics of quantum fields}, I.
Commun.  Math.  Phys.  {\bf 126} (1989) 217--247, \&
II. {\bf 130} (1990) 285--309.

\bibitem{L2}
R. Longo, {\it A duality for Hopf algebras and for subfactors},
Commun. Math. Phys. {\bf 159} (1994) 133--150.

\bibitem{L4}
R. Longo,
{\it Conformal subnets and intermediate subfactors}, 
to appear in Commun. Math. Phys., math.OA/0102196.

\bibitem{LR}
R. Longo \& K.-H. Rehren, {\it Nets of subfactors}, Rev. Math. Phys.
{\bf 7} (1995) 567--597.

\bibitem{LRo}
R. Longo \& J. E. Roberts, {\it A theory of dimension}, $K$-theory
{\bf 11} (1997) 103--159.

\bibitem{O1}
A. Ocneanu,
{\it Quantized group, string algebras and Galois theory for algebras},
in {\em Operator algebras and applications, Vol. 2 (Warwick, 1987)},
(ed. D. E.  Evans and M. Takesaki), London Mathematical Society
Lecture Note Series {\bf 36}, 
Cambridge University Press, Cambridge, 1988, 119--172.

\bibitem{O2}
A. Ocneanu,
{\it Paths on Coxeter diagrams:
from Platonic solids and singularities to minimal models and subfactors},
(Notes recorded by S. Goto), in {\em Lectures on operator theory},
(ed. B. V. Rajarama Bhat et al.),
The Fields Institute Monographs, AMS Publications, 2000, 243--323.

\bibitem{PP}
M. Pimsner \& S. Popa, {\it Entropy and index for subfactors},
Ann. Scient. \'{E}co. Norm. Sup. {\bf 19} (1986) 57--106.

\bibitem{P1}
S. Popa, 
{\it Orthogonal pairs of $*$-subalgebras in finite von Neumann algebras},
J. Operator Theory {\bf 9} (1983) 253--268.

\bibitem{P2}
S. Popa, {\it Symmetric enveloping algebras, amenability and AFD
properties for subfactors}, Math. Res. Lett.
{\bf 1} (1994) 409--425.

\bibitem{PS} A. Pressley \& G. Segal, `` Loop Groups'',
Oxford University Press, Oxford, 1986.

\bibitem{R1}
K.-H.  Rehren,
{\it Braid group statistics and their superselection rules}, 
in: ``The Algebraic Theory of Superselection Sectors'', D.  Kastler ed.,
World Scientific, Singapore, 1990.

\bibitem{R2}
K.-H. Rehren,
{\it Space-time fields and exchange fields},
Commun. Math. Phys. {\bf 132} (1990) 461--483.

\bibitem{R3}
K.-H. Rehren,
{\it A new view of the Virasoro algebra},
Lett. Math. Phys. {\bf 30} (1994) 125--130.

\bibitem{R4}
K.-H. Rehren, 
{\it Chiral observables and modular invariants},
Commun. Math. Phys.  {\bf 208} (2000) 689--712.

\bibitem{R5}
K.-H. Rehren, 
{\it Canonical tensor product subfactors},
Commun. Math. Phys. {\bf 211} (2000) 395--406.

\bibitem{R6}
K.-H. Rehren, 
{\it Locality and modular invariance in 2D conformal QFT},
in {\em Mathematical Physics in Mathematics and Physics} (ed. R. Longo),
Fields Inst. Commun. {\bf 30} (2001), AMS Publications, 341--354.
math-ph/0009004.

\bibitem{T}
M. Takesaki , ``Theory of Operator Algebras. I'' Springer-Verlag,
Berlin-Heidelberg-New York, 1979.

\bibitem{TL} V. Toledano Laredo,
{\it Fusion of positive energy representations of $LSpin_{2n}$},
Thesis, University of Cambridge, 1997.

\bibitem{Tu}  V. G. Turaev, ``Quantum Invariants of
Knots and 3-Manifolds'', Walter  de Gruyter, Berlin-New York, 1994.

\bibitem{W} A. Wassermann,
{\it Operator algebras and conformal field theory III: Fusion
of positive energy representations of $SU(N)$ using bounded operators},
Invent. Math. {\bf 133} (1998) 467--538.

\bibitem{X1}
F. Xu,
{\it New braided endomorphisms from conformal inclusions},
Commun. Math. Phys. {\bf 192} (1998) 347--403.

\bibitem{X2}
F. Xu,
{\it Applications of braided endomorphisms from conformal inclusions}, 
Internat.  Math. Res. Notices (1998) 5--23.

\bibitem{X3}
F. Xu,
{\it Jones-Wassermann subfactors for disconnected intervals},
Commun. Contemp. Math. {\bf 2} (2000) 307--347.

\bibitem{X4}
F. Xu,
{\it Algebraic coset conformal field theories I},
Commun. Math. Phys. {\bf 211} (2000) 1--44.

\bibitem{X5}
F. Xu,
{\it On a conjecture of Kac-Wakimoto},
Publ. RIMS, Kyoto Univ. {\bf 37} (2001) 165--190.

\bibitem{X6}
F. Xu,
{\it 3-manifold invariants from cosets}, preprint 1999,
math.GT/9907077.

\bibitem{X7}
F. Xu,
{\it Algebraic orbifold conformal field theories},
Proc. Nat. Acad. Sci. U.S.A.
{\bf 97} (2000) 14069--14073.

\end{thebibliography}}
\end{document}